\documentclass[pre,showpacs,twocolumn,superscriptaddress]{revtex4-1}
\usepackage{graphicx,amssymb,color}
\usepackage[dvipsnames]{xcolor}
 \usepackage{hyperref}

\begin{document}
\title{Universality and criticality of a second-order granular solid-liquid-like phase transition}
\author{Gustavo Castillo}
\affiliation{Departamento de F\'isica, Facultad de Ciencias F\'isicas y Matem\'aticas, Universidad de Chile, Avenida Blanco Encalada 2008, Santiago, Chile}
\affiliation{Laboratoire de Physique Statistique, Ecole Normale Sup\'erieure, UMR CNRS 8550, 24 Rue Lhomond, 75005 Paris, France}
\author{Nicol\'as Mujica}
\email[Corresponding author: ]{nmujica@dfi.uchile.cl} 
\author{Rodrigo Soto}
\affiliation{Departamento de F\'isica, Facultad de Ciencias F\'isicas y Matem\'aticas, Universidad de Chile, Avenida Blanco Encalada 2008, Santiago, Chile}
\date{\today}

\begin{abstract}

We experimentally study the critical properties of the non-equilibrium solid-liquid-like transition that takes place in vibrated granular matter. The critical dynamics is characterized by the coupling of the density field with the bond-orientational order parameter $Q_4$, which measures the degree of local crystallization. 
Two setups are compared, which present the transition at different critical accelerations as a a result of modifying the energy dissipation parameters. In both setups five independent critical exponents are measured, associated to different properties of $Q_4$: the correlation length, relaxation time, vanishing wavenumber limit (static susceptibility), the hydrodynamic regime of the pair correlation function, and the amplitude of the order parameter. The respective critical exponents agree in both setups and are given by $\nu_{\perp} = 1$, $\nu_{\parallel} = 2$, $\gamma = 1$, $\eta \approx 0.6 - 0.67$, and $\beta=1/2$, whereas the dynamical critical exponent is $z = \nu_{\parallel}/\nu_{\perp}  = 2$.
The agreement on five exponents is an exigent test for the universality of the transition. Thus, while dissipation is strictly necessary to form the crystal, the path the system undergoes towards the phase separation is part of a well defined universality class.
In fact, the local order shows critical properties while density does not. Being the later conserved, the appropriate model that couples both is model C in the Hohenberg and Halperin classification. The measured exponents are in accord with the non-equilibrium extension to model C if we assume that $\alpha$, the exponent associated in equilibrium to the specific heat divergence but with no counterpart in this non-equilibrium experiment, vanishes.

\end{abstract}
\maketitle

\section{Introduction}

Thermodynamics and equilibrium statistical mechanics have shown an extraordinary success in  describing phase transitions. It has been possible to classify them, predict their properties, and understand the critical phenomena. Also, they provide a framework that can be used to build efficient numerical tools as the Monte Carlo simulations or to design and analyze experiments, for example exploiting the concept of universality classes. From a theoretical viewpoint the critical phase transitions were classified in universality classes in terms of symmetry and conservation properties \cite{BookCritical,hohenberg}.

In several  order-disorder phase transitions the dynamics of the transition needs the interplay of two or more order parameters \cite{Hoh,TanakaNature,vanThiel,Gorentsveig,Glosli,Shimizu,Kockelkoren}. One particularly interesting case is the so called model C in the Hohenberg and Halperin classification, where a non-conserved critical order parameter couples to the conserved noncritical density \cite{hohenberg,Kockelkoren,Ccorrect}. Examples where this model can be applied are varied, comprising   liquid-liquid phase transitions \cite{vanThiel,Tanakaliquid,Glosli,Shimizu}, binary alloys \cite{Gorentsveig}, and anisotropic magnets \cite{Dudka}. 

In out of equilibrium conditions there is no such systematic framework that can be used to analyze or classify phase transitions. There are, nevertheless, some notorious examples of prototype models that have been shown to be quite general, allowing other systems to be compared with them. Some of these are, for example, the directed percolation process, the Kardar-Parisi-Zhang model for surface growth and the Swift-Hohenberg model \cite{MarroNonEq,HenkelNonEq,KPZ,Takeuchi2014,CrossHohenbergRMP}. 
The critical phenomena methodology in equilibrium has been extended to these cases, where again we find critical exponents, universality classes, and scaling functions, with the dynamic renormalization group as a useful approach. But, still much understanding is needed to reach a state of knowledge comparable to the equilibrium case.

Granular matter, due to the strong dissipation present at the particle interactions and the corresponding need for continuous energy injection to sustain dynamic states, is an excellent candidate for studying out of equilibrium phase transitions. Important progress has been made along this direction \cite{Melby2005, prevost2004, Reis2006, cartes, clerc2008,Reis2002,YFan, CMay,Pacheco,tanaka2008}, but a wider view in the context of dynamical phase transitions is still lacking. Recently, we presented an experimental study of a granular liquid-solid-like phase transition in a vibrated quasi-two-dimensional granular system \cite{castillo_PRL}. There, we showed that the transition is characterized not only by the density field but also by a bond-orientational order parameter, which is described by the model C. To our understanding, this is the first non-equilibrium realization of this class of phase transition. 

In this manuscript we extend the results found previously by focusing now on the universality of the transition. That is, we aim to verify if by varying some experimental parameters, the same critical exponents are found. We remark that in Ref. \cite{castillo_PRL} five independent critical exponents were found which, therefore, put an exigent  test the universality condition.

The article is organized as follows. Section \ref{sec.Q2D} describes the liquid-solid-like transition that takes place in confined quasi two dimensional granular systems. In Sec. \ref{sec.experiments} the experimental setup and procedures are explained, describing the two configurations that are used to test the universality of the critical exponents. The experimental results and the determination of the critical accelerations and exponents are presented in Sec. \ref{sec.results}. Finally, conclusions are given in Sec. \ref{sec.conclusions}.

\section{The liquid-solid-like transition in quasi two dimensional granular systems} \label{sec.Q2D}

Granular matter is ubiquitous in our daily life, still a fully understanding of its dynamical behavior has remained rather elusive for several years \cite{Aranson, Jaeger}. Dry granular matter is a collection of athermal macroscopic particles that interact mainly through hard core-like dissipative collisions, and depending on external conditions such as pressure or packing fraction it may behave as a solid, a liquid or even a gas. 

Energy injection is needed to compensate the energy dissipated in grain-grain and grain-wall collisions. Vibrations are an efficient method to perform this task in a distributed and controlled way, being possible to reach stable states. Among the different possibilities to perform the vibrations, the quasi-two dimensional (Q2D) geometry has gained attention because it allows the granular system to be followed in detail at the individual grain scale together with the collective dynamics. In Q2D systems, grains are placed in shallow box, with a height that is between one and two particle diameters, while it is large in the horizontal directions. The box is vertically vibrated with maximal accelerations larger than gravity such that grains acquire vertical energy that is transferred via collisions to the horizontal degrees of freedom. In the reduced two-dimensional dynamics, the system resembles a liquid but with the important difference that the collisions are not conservative.

For fixed density and vibration frequency, by increasing the maximum acceleration the system presents a phase transition where grains cluster in solid-like regions with crystalline order \cite{prevost2004,Melby2005}. The instability is originated in the development of an effective negative compressibility \cite{Argentina2002,clerc2008} and the transient dynamics is governed by waves \cite{clerc2008}. Once the system is segregated the solid-like cluster show fluctuations in the interface that are well described by the capillary theory, allowing us to extract an effective surface tension that has non-equilibrium origin \cite{Luu_PRE}.

In our previous work we studied experimentally the solid-liquid-like phase transition in the vibrated Q2D geometry. Depending on the filling fraction and height of the cell, the transition was either discontinuous or continuous. In the later case critical phenomena develop, with static and dynamical properties that show power law behavior in terms of the distance to the critical acceleration \cite{castillo_PRL}.

\section{Experimental setup and procedures} \label{sec.experiments}

\begin{figure}[t!]
\begin{center}
\includegraphics[width=\columnwidth]{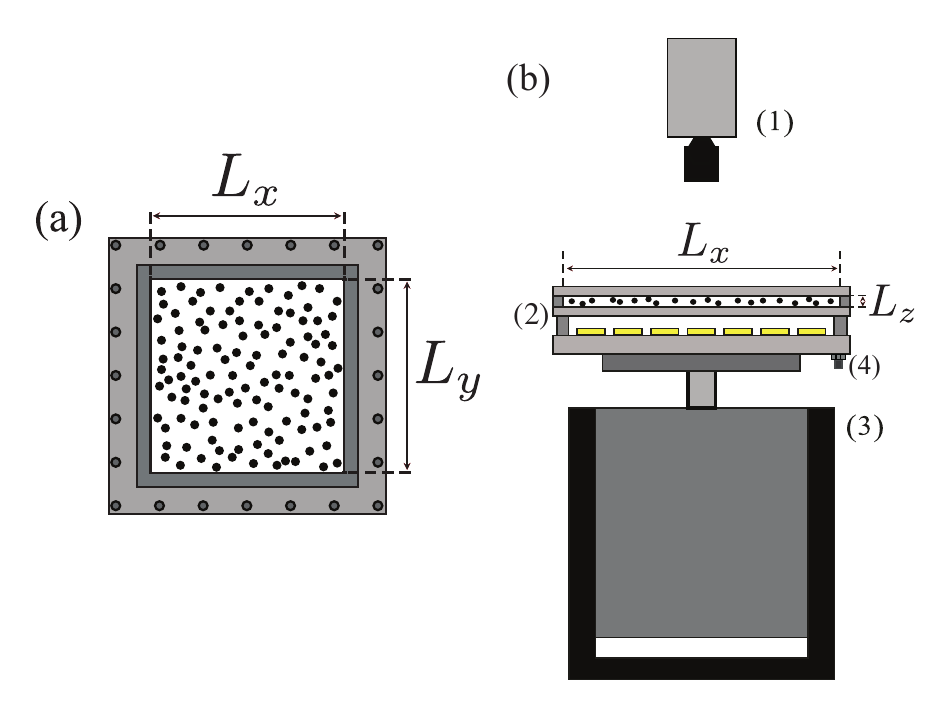}
\caption{Schematic of the experimental setup. (a) Top view of the quasi-2D cell, with $L_x = L_y = 100$d. (b) Side view of the setup. The vertical height in the cell is $L_z = 1.94d\pm 0.02d$. The cell is illuminated from below with a 2D array of light emitting diodes, which light is diffused with a white acrylic sheet placed between the array and the cell. (1) camera, (2) quasi-2D cell, (3) electromechanical shaker, (4)~accelerometer. }
\label{fig1}
\end{center}
\end{figure}

In this paper we extend the analysis of the previously published study \cite{castillo_PRL}. Two sets of experiments  are used to test the universality hypothesis. They differ in the dissipation coefficients, which will be labeled experiments A and B, with larger and lower dissipation respectively. 

\begin{figure}[t!]
\begin{center}
\includegraphics[width=\columnwidth]{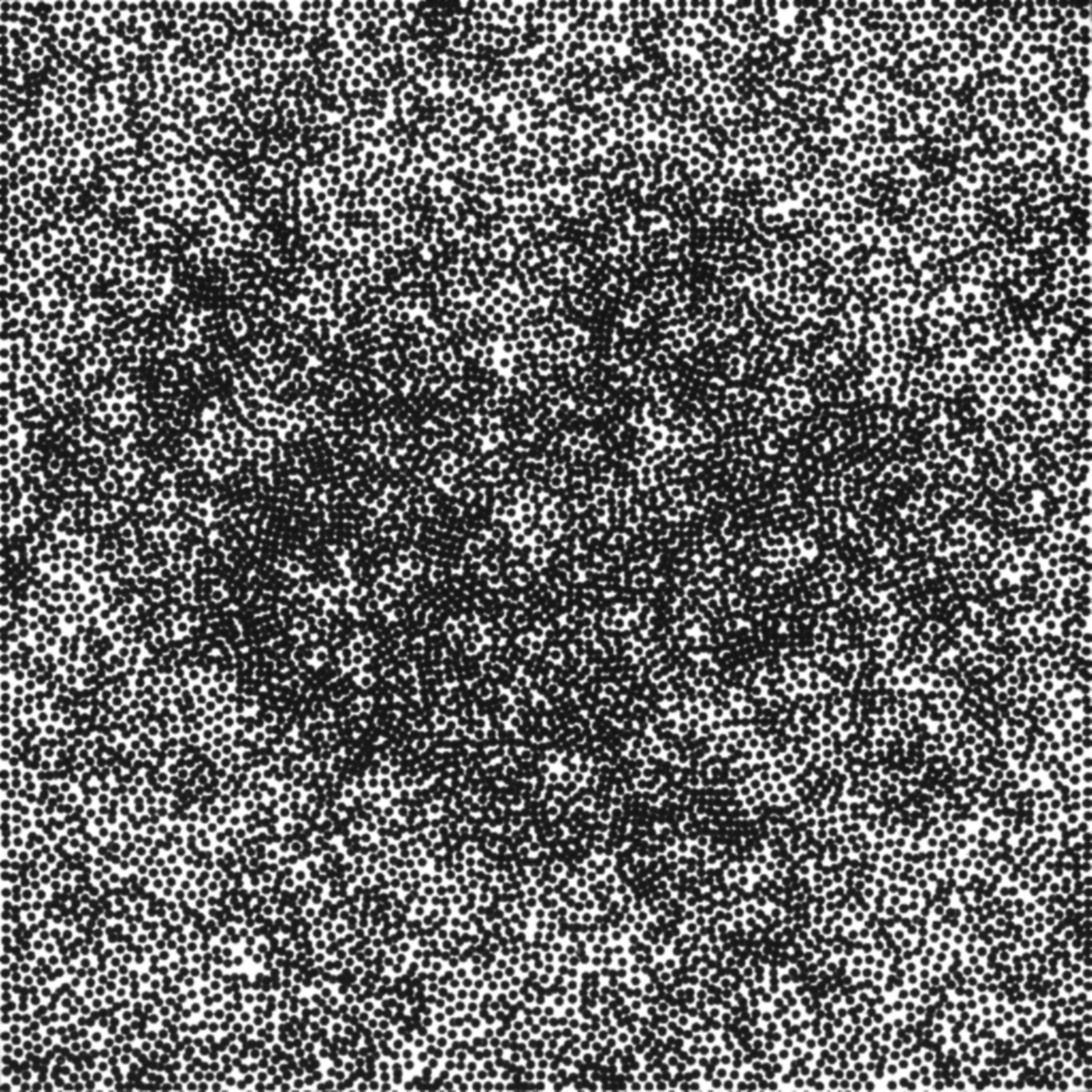}
\caption{A typical raw image of the complete system, for type B experiment and $\Gamma =  4.05 \pm 0.01$. Particles are visualized as black disks on a white background.}
\label{raw_img}
\end{center}
\end{figure}

The granular system is composed of $N \sim 10^4$ stainless steel spherical particles of mass $m=4.45\times 10^{-3}\mbox{ g}$, and  $d=1$ mm diameter. The Q2D box has lateral dimensions $L_x = L_y \equiv L = 100d$. The box consists of two $10$~mm thick glass plates separated by a square metallic frame. Each inner glass surface has an indium tin oxide (ITO) coating, which dissipates electrostatic charges generated by collisions of particles with the walls. The box is fixed to a base by four posts placed at each corner of the cell. The base supports an array of high intensity light emitting diodes. A piezoelectric accelerometer is fixed to the base, allowing the measurement of the imposed forcing acceleration with a resolution of $0.01$g. The main advantage of this setup is that particles are illuminated from below. They are then visualized as dark disks on a white background. This allows to detect particles in dense clusters, even when particles are partially mounted on top of each other. A scheme of the setup is shown in Fig.\ \ref{fig1}. A typical image of the system is shown in Fig. \ref{raw_img}.

The whole setup is forced sinusoidally with an electromechanical shaker, with vertical displacement $z(t) = A \sin(\omega t)$. Top view images are obtained with a camera at $10$ fps. The images acquired have a typical resolution of $1020\times1020 \mbox{ pix}^2$, thus we obtain particles of 10 pix diameter approximately. Particle positions are determined at sub-pixel accuracy.  Results have been obtained by fixing the particle number $N$, cell height $L_z$ and driving frequency $f=\omega/2\pi = 1/T=80$ Hz. The dimensionless acceleration $\Gamma = A\omega^2/g$ is varied in the range $1-6$. 
In this present work we focus on the configuration C2 of reference  \cite{castillo_PRL}, namely $L_z = 1.94d \pm 0.02d$ and $N = 11504$, which implies the filling fraction to be $\phi = N\pi d^2/4L^2=0.904$. This filling fraction value is possible because the system is Q2D, which allows to accommodate more particles than for a pure 2D system.

As was demonstrated in \cite{castillo_PRL}, under this particular configuration the system undergoes a second-order type phase transition. The aim of this paper is to present a more detailed study of the transition itself. Additionally, we are particularly interested if there is universality in this granular out-of-equilibrium phase transition. In order to study the universal behavior of the critical exponents, we change the dissipation parameters of the system by changing both the bottom and top lids. We use two pairs of ITO (indium tin oxide) coated glass plates. In the experiments of type A the coating is 25 nm thick, whereas for type B experiments it is 750 nm thick. It has been shown that ITO films mechanical, fracture, morphological and electric properties depend on fabrication details such as thickness, annealing, substrate, feed gas, among others \cite{Leterrier_2004}. So, it is expected that the dissipation parameters, namely inelastic and friction coefficients, should be different for the two sets of ITO coated glass plates. As will be shown below, our results indeed demonstrate that the dissipation is stronger for the thinner ITO coated glass plates (type A experiments).

\begin{figure*}[t!]
\begin{center}
\includegraphics[width=0.68\columnwidth]{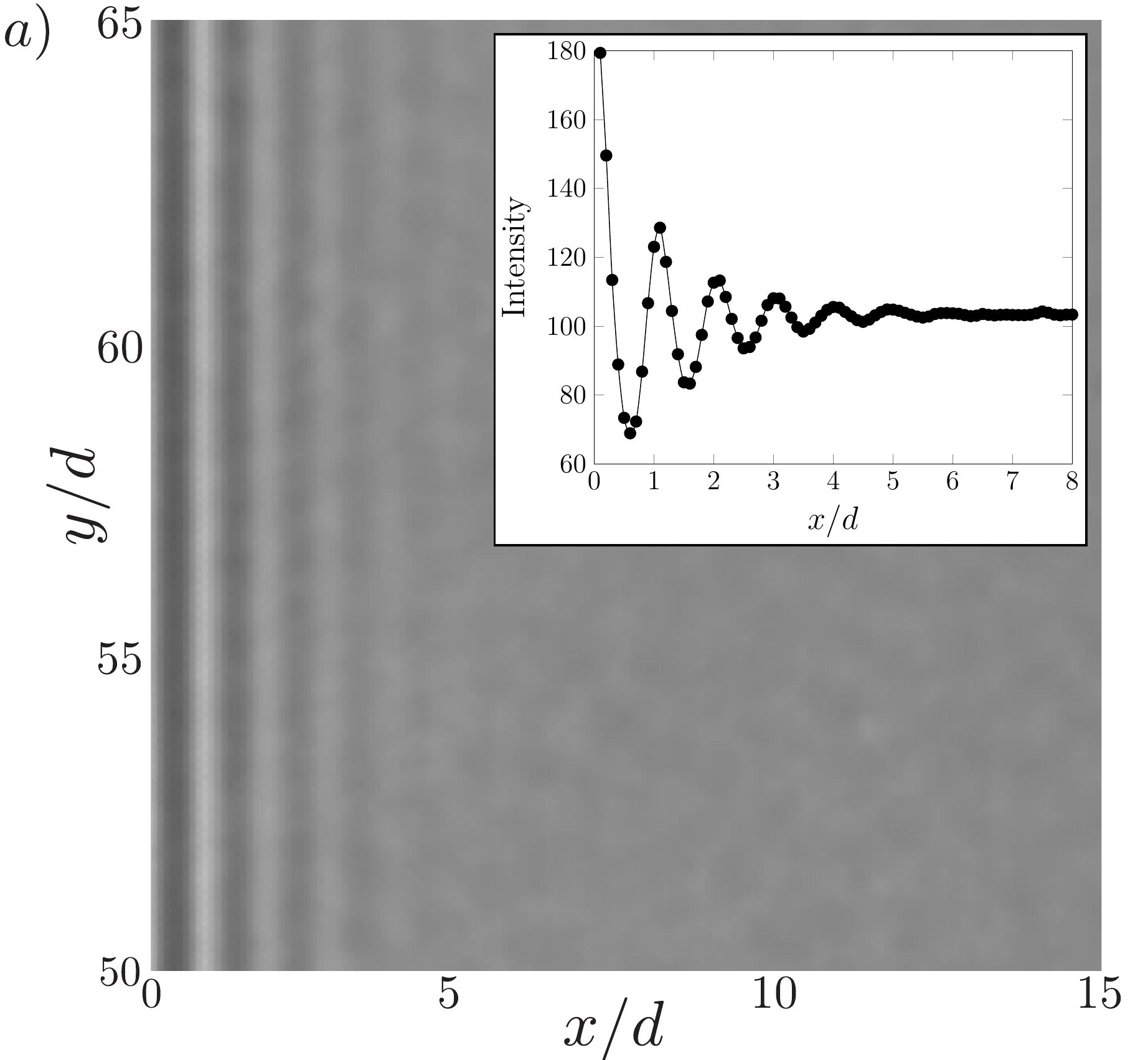}
\includegraphics[width=0.68\columnwidth]{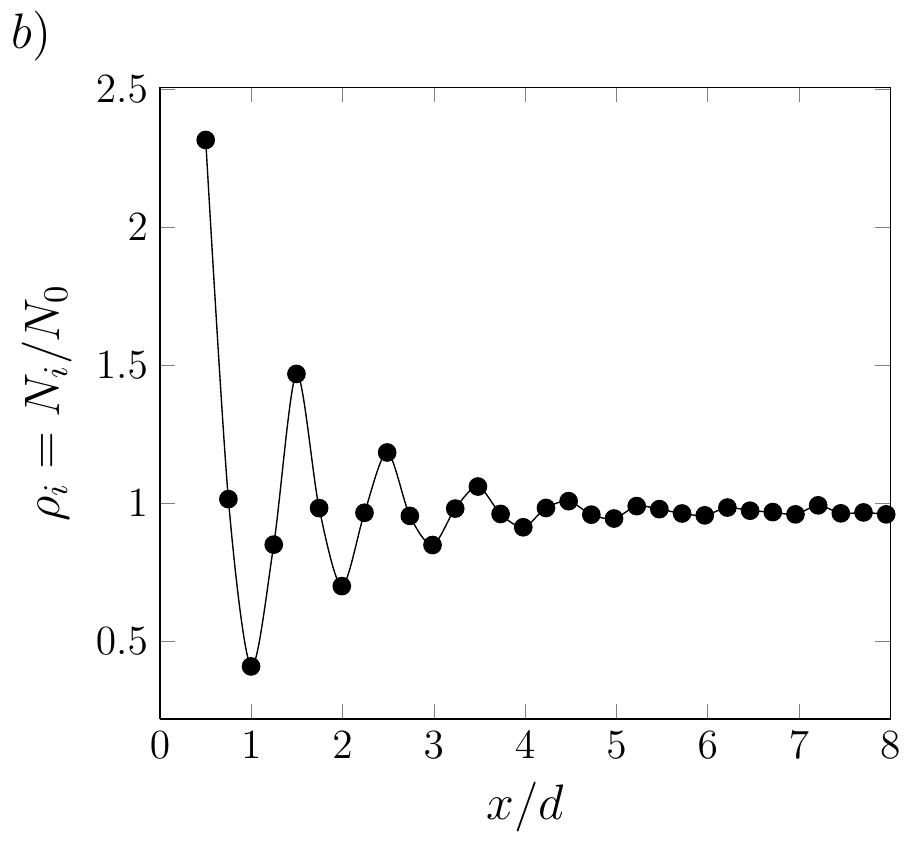}
\includegraphics[width=0.68\columnwidth]{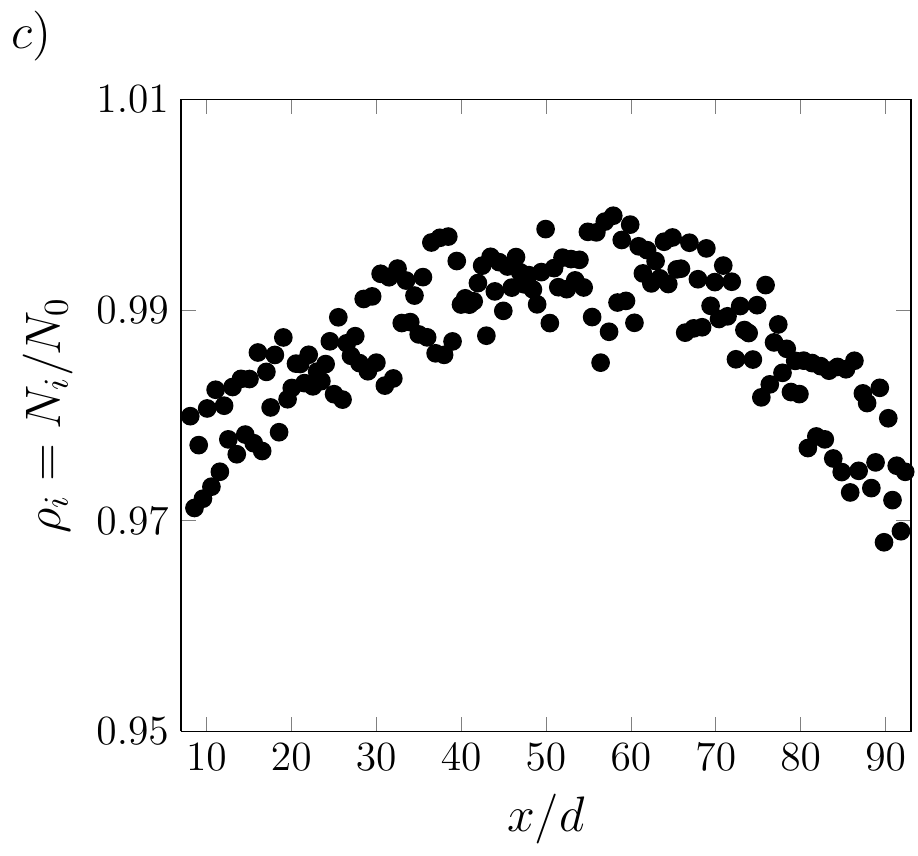}
\caption{(a) Average image obtained from 3270 raw images.  Only a section of the setup near a wall is shown (of size $15d\times15d$). Regions in dark grey (low intensity) results from more particles in these zones. The vertical stripes at the left account for layers of particles that move less and spend more time near the wall because of the extra dissipation at the side wall. The inset shows the intensity of this image as a function of position, showing that for $x \gtrsim 5d$ the system is almost homogeneous. (b)~Time averaged coarse-grained particle density obtained directly from particle detection as a function of $x$, with $d/4$ as bin width. It also shows oscillations due to side wall layering and confirms the homogeneity for $x \gtrsim 5d$ at this scale. (c) The same coarse-grained density at a different scale, 
showing that the 
density at the center is about $2\%$ higher than close to the edges.
Here the bin width is $d/2$. For both (b) and (c) $N_i$ is the average number of particles detected in a vertical bin spanning $y = [0,100d]$ and $x_i = [(i-1)w_b, iw_b)$, where $w_b = d/4$ or $d/2$ is the bin width. $N_{0}$ is defined as the number of particles that would fit into a bin with an homogeneous density $\rho_o = N/L^2$.}
\label{fig3}
\end{center}
\end{figure*}

The ITO coating works very well for many hours of experimental runs. Eventually, it does however get damaged by particle collisions. The precise time of reproducible runs is probably function of the ITO coating thickness. This is supported by the thickness dependence of the crack onset strain in ITO thin films, which is lower for thicker coatings \cite{Leterrier_2004}. All data presented in this paper correspond to reproducible runs for which no important damage was noticeable. In fact, a surface damaged ITO coating is manifested in important changes of the measured quantities -- such as the density and order structure factors -- with respect to those obtained for a new pair of ITO coated glass plates. We conjecture that the damage occurs because of erosion of the ITO coating, which in turn affects particle interactions by an increase of electrostatic forces and contamination of the system by dust formed from the ITO coating.
In order to insure reproducibility, glass plates were changed periodically during the time duration of all the experimental runs, and all parts of the experiments are cleaned in an ultrasonic bath before mounting the experimental cell again, including the particles. 

Two important issues are the top and bottom plates flatness and homogeneity of $L_z$ through the experimental cell. With the cell empty of particles the height is measured at nine different positions on a regular spaced grid with an optical microscope (see supplementary material of \cite{castillo_PRL}). Within experimental errors the vertical height is homogeneous, $L_z = 1.94d \pm 0.02d$. However, the homogeneity of $L_z$ does not insure that the glass plates are both flat. Because of the mechanical constrains and stresses exerted on the plates, some small residual curvature could exist. 

In Fig. \ref{fig3}(a) we show an average of 3270 images for $\Gamma=2.05$, well below the solid-liquid transition (for type B experiments, $\Gamma_c \approx 4.5$). Only a small part of the setup is shown, of size $15d\times15d$. Viewed from above it corresponds to the left side wall at almost half of the cell. It is clear that near the side wall particles tend to form layers of thickness $\sim d$, within which particles barely move compared with the rest of the system. This layering, observed at all side walls, is induced by the extra dissipation given by the side wall collisions, similar to what has been shown in sheared granular matter \cite{YFan}. However, for a distance $\gtrsim 5d$ from a side wall, the system is nearly homogeneous. The inset of Fig. \ref{fig3}(a) shows an average intensity plot as a function of the position (time and $y$ averaged), showing the layers close to the edge and then saturating at a constant value. Also, by means of a time average coarse-graining procedure we can compute the density of particles as a function of distance from a wall. This is shown in Figs. \ref{fig3}(b) and \ref{fig3}(c). Close to the wall we notice the same particle layering and an apparent homogenization for $x\gtrsim 5d$ (Fig. \ref{fig3}(b)). The asymptotic limit is slightly less than expected to compensate for the particle excess at the side walls. At a larger scale, as shown in Fig. \ref{fig3}(c), we observe that there is a small density gradient that leads to an excess of particles at the cell's center compared to the edges: density is about $2\%$ higher at the center that near the side walls. This can be the result of three effects, which are probably all present. Firstly, even though the cell's height is homogeneous within experimental errors, there might be a small residual concavity, which makes the particles to accumulate near the center. A second factor could be that the vertical acceleration may not be constant throughout the cell. In other words, there could be some ``flapping" of the system. Indeed, we have measured the acceleration of the cell at different positions and we find that it is about $0.2\%$ higher near the borders that at the center. Finally, the presence of dissipative side walls could induce such large scale inhomogeneity.

Another important issue is the mechanical leveling of the whole setup (for details, see supplementary material of Ref. \cite{castillo_PRL}). The cell's horizontality, and thus the system's isotropy, can be checked by two ways: First, it is verified through the static structure factor $S(\vec k)$ (defined below) and the average density Fourier components $\langle \rho({\vec k,t}) \rangle$ (for now we can say that $S(\vec k)$ is a measure of density fluctuations in Fourier space). When these quantities are plotted versus $k_x$ and $k_y$ there is no preferential direction. For example, $S(k_x,k_y)$ shows the characteristic symmetric circular annular shape at $kd = 2\pi$, where $k = |\vec k|$. A second verification is given by the even symmetry of the coarse-grained density with respect to the cell center line, as can be observed from the data presented in Fig. \ref{fig3}(c), which is indeed symmetric with respect to $x/d = 50$.

The particle detection is done by using a modified open source Matlab code which uses a least-square algorithm \cite{Shattuck}. Our modified version in C++  \& CUDA allows faster computation for large number of particles \cite{memoriaJuan, juansilva}. The algorithm allows to detect both layers of particles in a dense solid cluster, where the top layer particles are placed in the valleys that the bottom particles form. Typical experimental runs consist of at least 30 video acquisitions, one for each $A$, of about 3300 images each. Therefore, the complete number of images to analyze for a single experimental run is about $10^5$.

Finally, for fixed geometry, particle density and vibration frequency we perform vibration amplitude ramps, from $\Gamma \gtrsim 1$ in the liquid phase, increasing $A$ going through the solid-liquid transition that is reached at $\Gamma = \Gamma_c$. In order to check that the transition is continuous, for some runs we also perform decreasing amplitude ramps starting above $\Gamma_c$.

\begin{figure*}[t!]
\begin{center}

\includegraphics[width=\columnwidth]{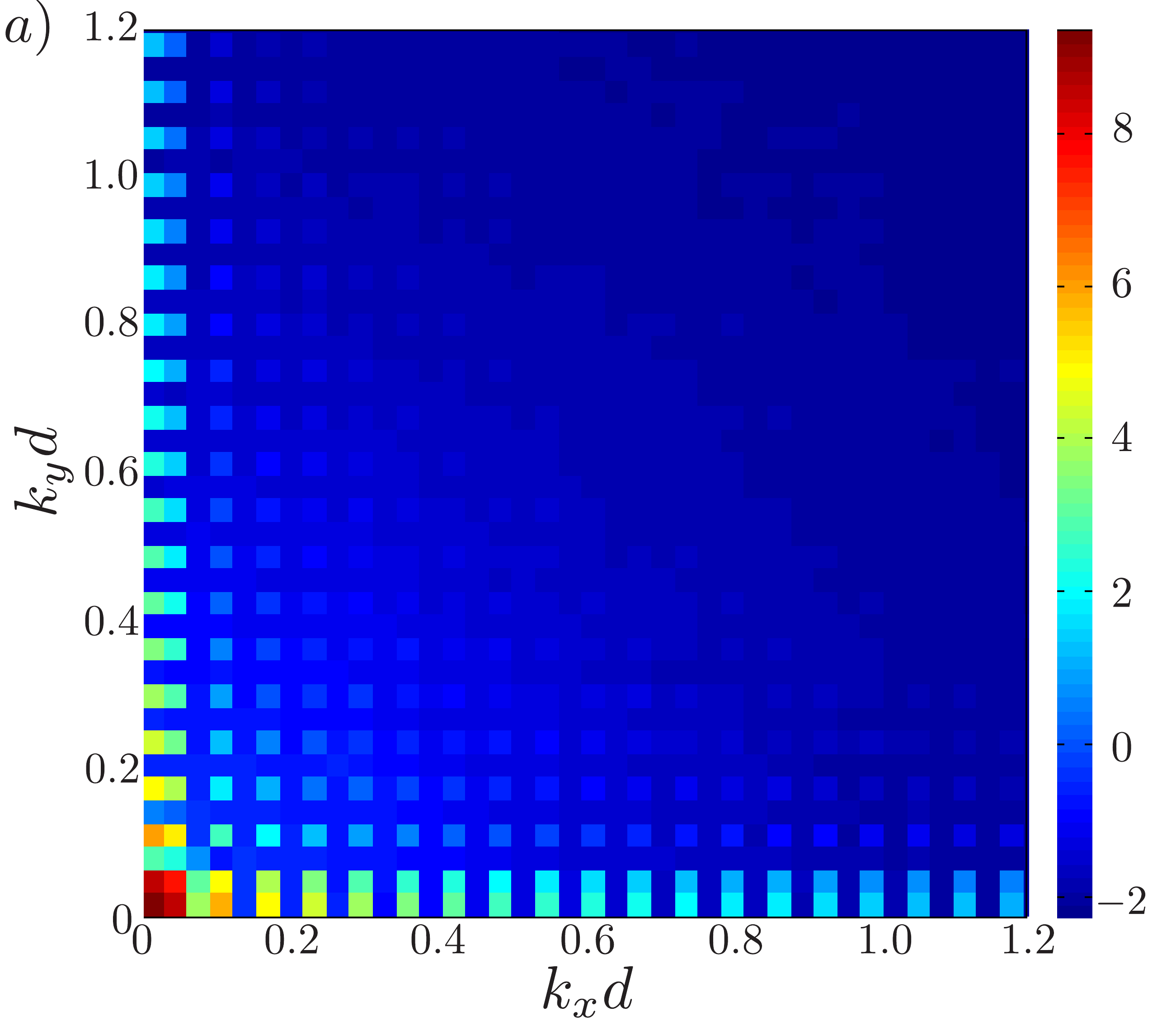}
\includegraphics[width=\columnwidth]{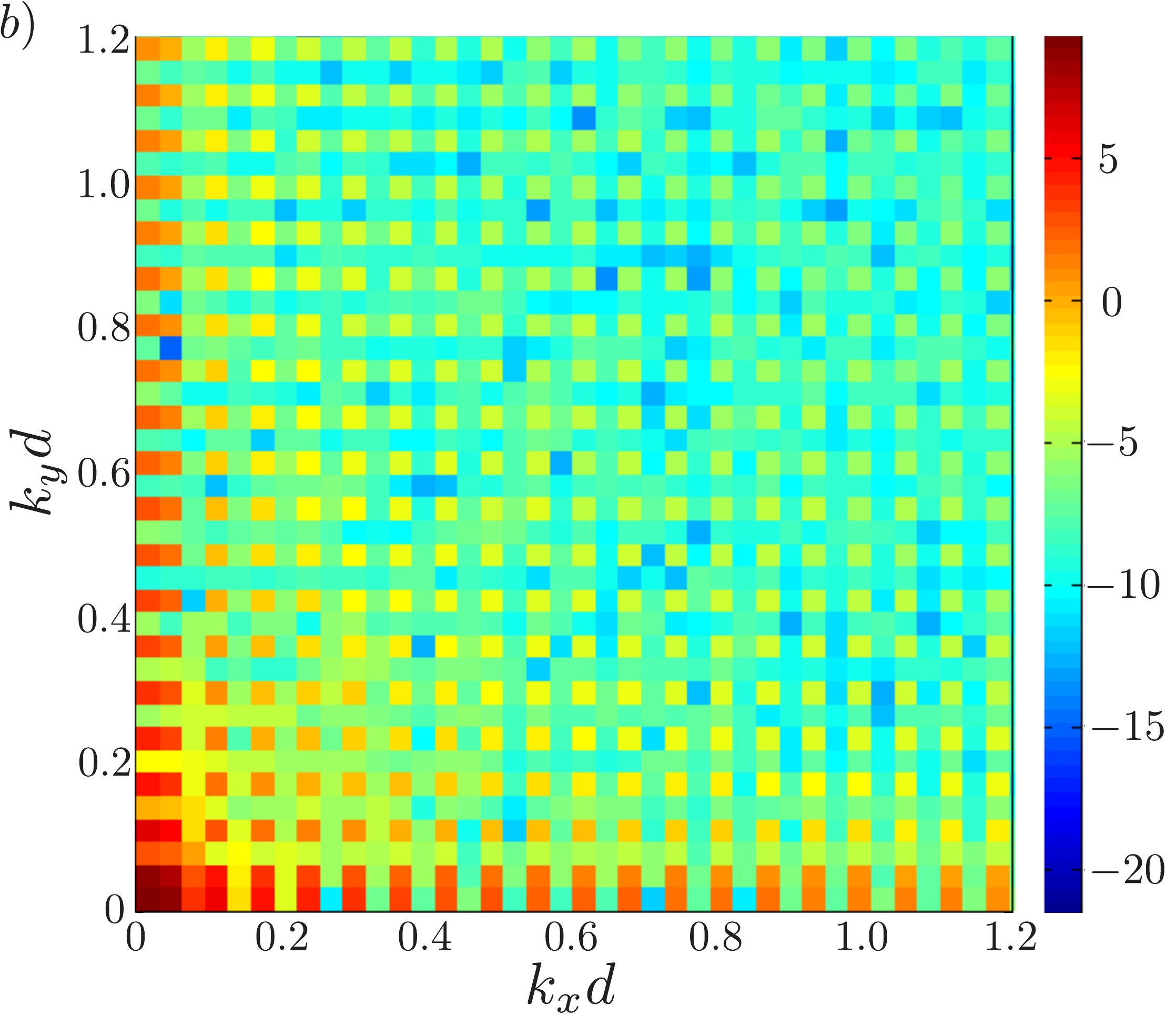}
\includegraphics[width=\columnwidth]{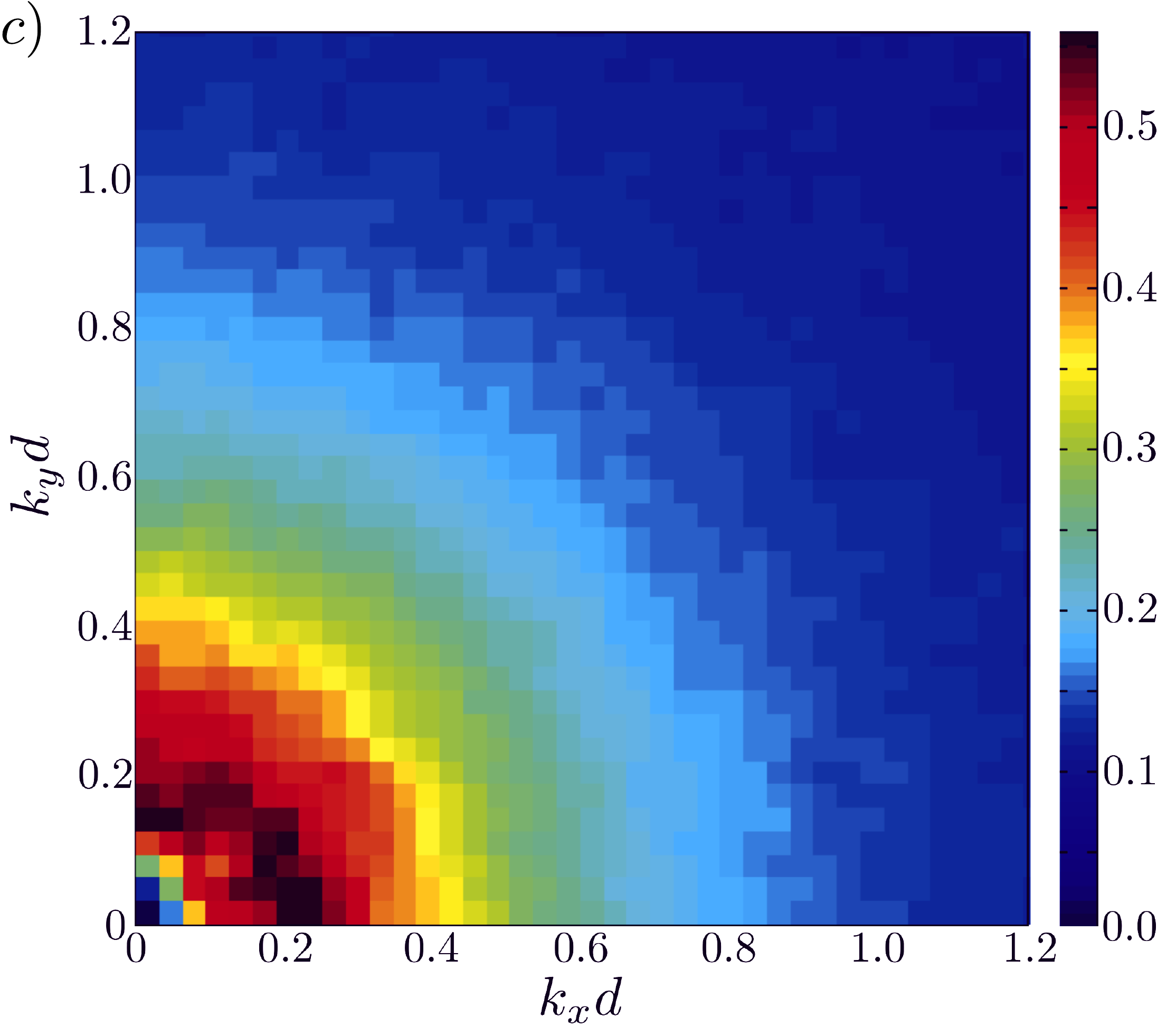}
\includegraphics[width=0.95\columnwidth]{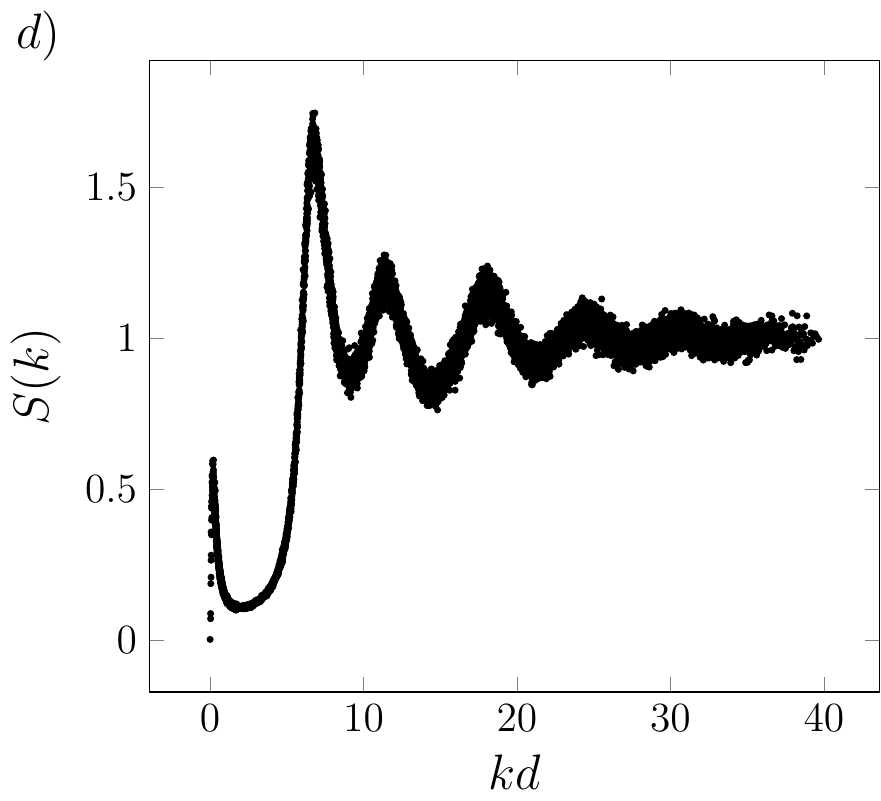}
\caption{(Color online) Color plots of $\log_{10}{(\langle \rho(\vec k,t)\rho(\vec k,t)^*\rangle/N )}$ (a), $\log_{10}{(\langle \rho(\vec k,t)\rangle \langle \rho(\vec k,t)^*\rangle/N )}$  (b) and the structure factor function $S(\vec k)$ (c) as functions of $k_x$ and $k_y$. From $S(\vec k)$ the isotropy of the fluctuations is clear from the circular halo shape of the pre-peak. (d) $S(k)$ for a wide range of $kd$, obtained from $S(k_x,k_y)$ by simply plotting as a function of $k = \sqrt{k_x^2 + k_y^2}$. For all these figures $\Gamma=3.75<\Gamma_c$ and data was obtained from type B experiment.}
\label{fig4}
\end{center}
\end{figure*}

\begin{figure*}[t!]
\begin{center}
\includegraphics[width=0.96\columnwidth]{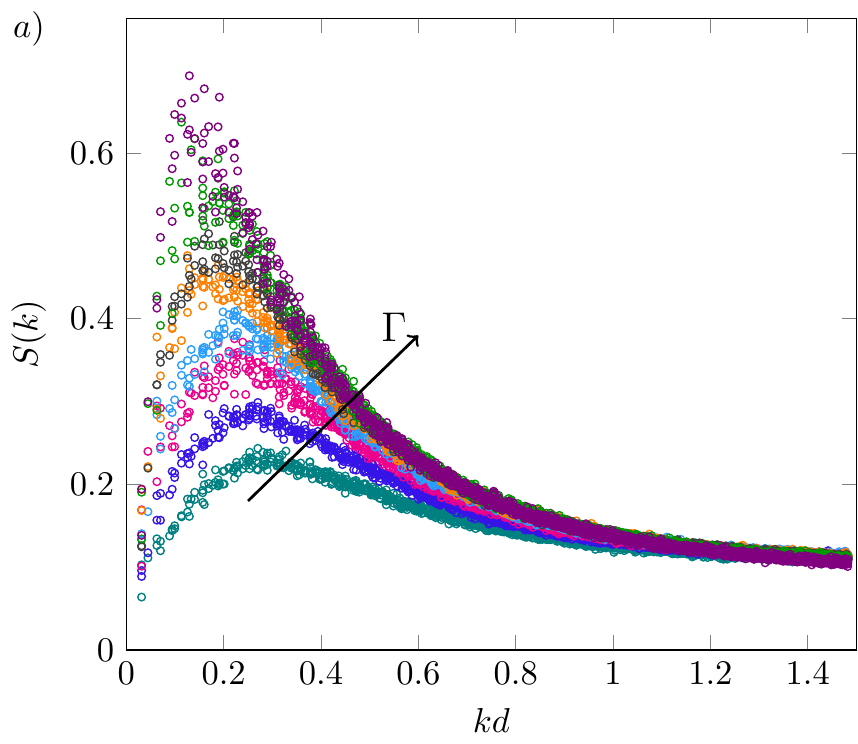}\qquad
\includegraphics[width=0.96\columnwidth]{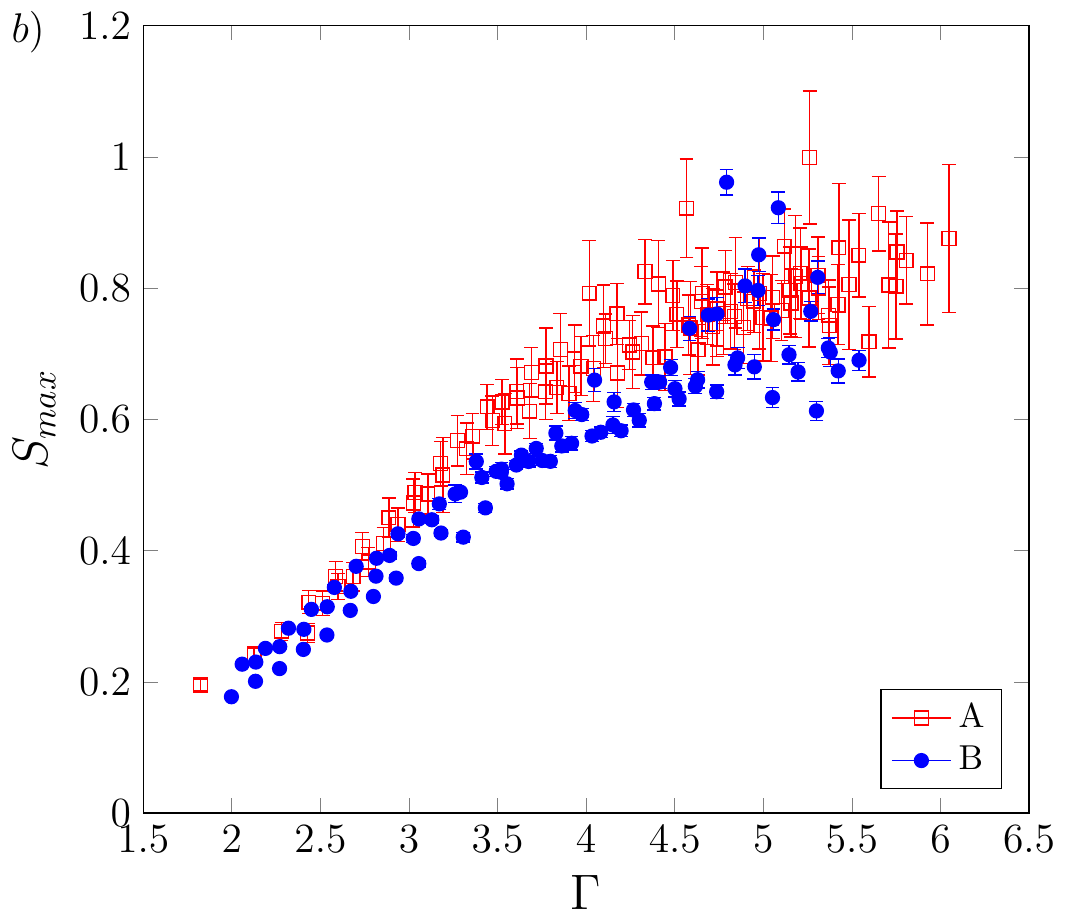} 
\vspace{0.7cm}

\includegraphics[width=\columnwidth]{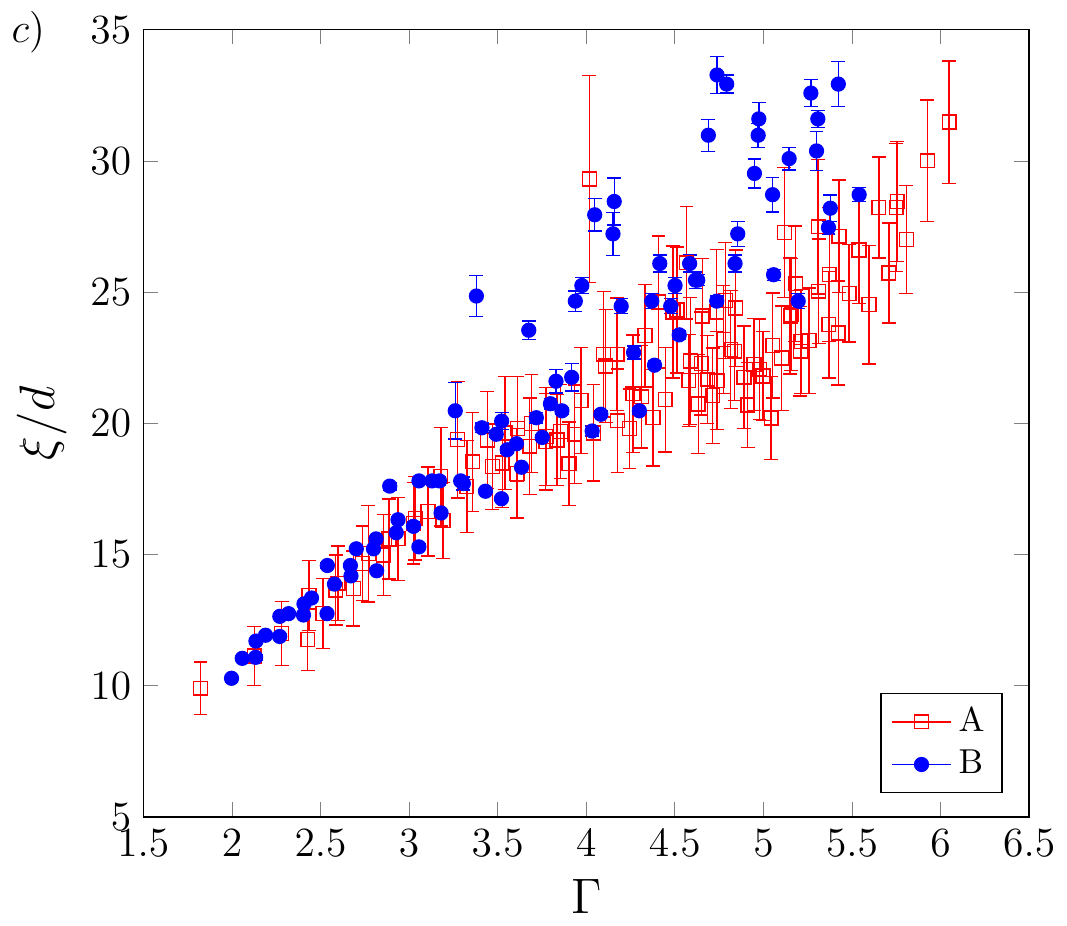}
\includegraphics[width=0.95\columnwidth]{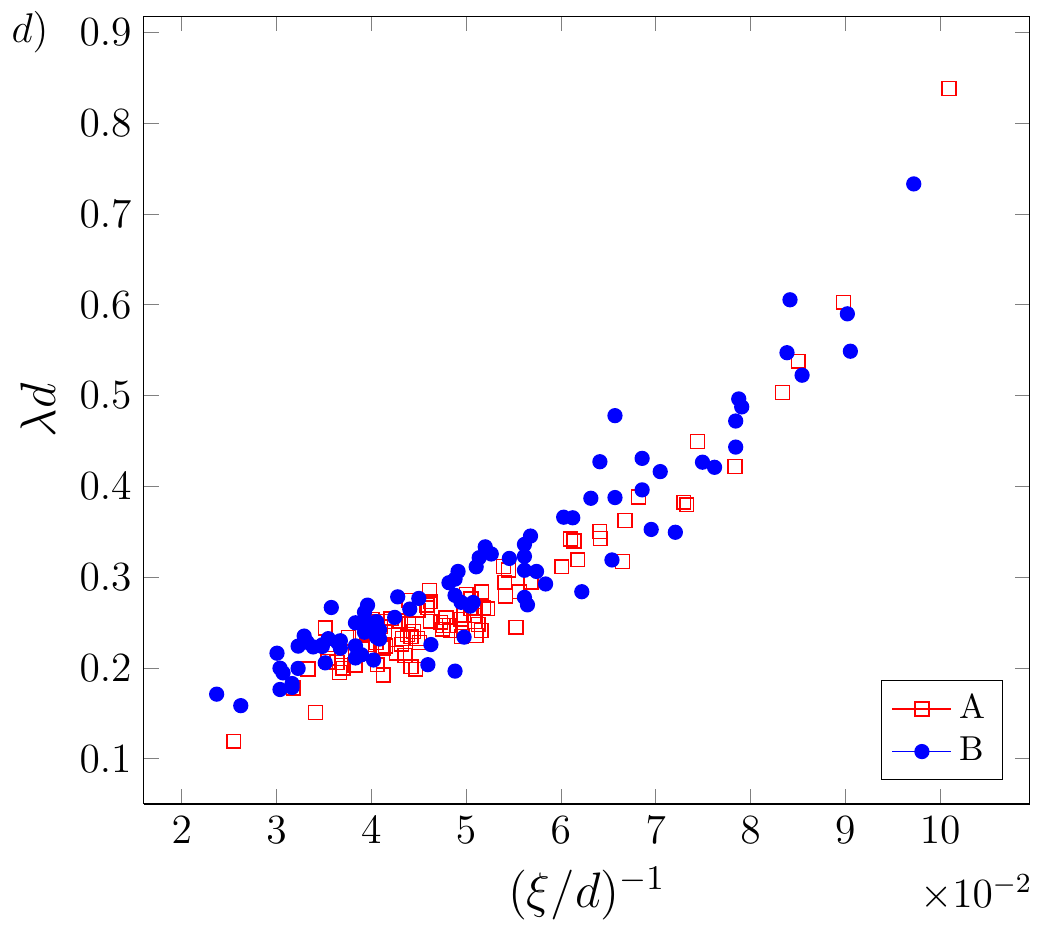}
\caption{(Color online) (a) $S(k)$ in the large wavelength limit for different accelerations $\Gamma$. The pre-peak grows and shifts towards lower $k$ for increasing $\Gamma$. We define the pre-peak's maximum $S_{\rm max}$ that is obtained at $k=k^*$. The width $\lambda$ of the pre-peak is defined as the width at half height. (b,c) $S_{\rm max}$ and $\xi/d = \pi/k^*$ as functions of $\Gamma$ for both types of experiments. (d) $\lambda d$ versus $d/\xi$. A linear dependence is observed for $\xi \gtrsim 15d$ ($d/\xi \lesssim 0.07$).  }
\label{fig5}
\end{center}
\end{figure*}

\section{Experimental results} \label{sec.results}

\subsection{Static structure function.}
Particle positions $\vec r_j(t)$ in the plane $(x,y)$ are determined for each time $t$.
Experimentally, there is no access to the $z$ coordinate. Thus, the 2D microscopic density field Fourier components are 
\begin{eqnarray}
{\rho}(\vec k,t) &=& \int \mbox{d}^2\vec{r}\, e^{i\vec r\cdot\vec k} \rho(\vec r,t)=\sum_{j=1}^{N} e^{i \vec k \cdot \vec r_j(t)}.
\end{eqnarray}
The static structure factor $S(\vec k)$ measures the intensity of density fluctuations in Fourier space:
\begin{eqnarray}
S(\vec k) &=& \frac{\langle|\rho(\vec k,t) - \langle \rho(\vec k,t) \rangle|^2\rangle}{N},\\
S(\vec k) &=&\frac{\langle \rho(\vec k,t) \rho(\vec k,t)^*\rangle-\langle \rho(\vec k,t)\rangle \langle \rho(\vec k,t)^*\rangle}{N}, \label{defSk}
\end{eqnarray}
where $\langle \,\, \rangle$ denotes time averaging. In general $\langle \rho(\vec k,t) \rangle~\neq~0$, due to inhomogeneities induced by boundary conditions, as those shown in Fig. \ref{fig3}. The wave vectors are computed from $\vec k = \pi ( n_x \hat {\i} +  n_y\hat  {\j})/L$, where $n_x, n_y \in \mathbb{N}$, and $k = |\vec k|$.

In Figs. \ref{fig4}(a) and \ref{fig4}(b) we present color plots of the two terms that are used for the computation of $S(k_x,k_y)$, namely $\langle \rho(\vec k,t) \rho(\vec k,t)^*\rangle/N$ and $\langle \rho(\vec k,t)\rangle \langle \rho(\vec k,t)^*\rangle/N$ as functions of $k_x$ and $k_y$ (both in $\log_{10}$ scale). It turns out that both quantities are strongly non-monotonic and are different by several orders of magnitude. In the lower wavenumber range that is plotted these quantities show a set of peaks placed on a regular grid on top of a smooth background, taking large values when both $n_x$ and $n_y$ are odd, whereas the other modes are much lower. This is understood by the even symmetry that the density field has with respect to the cell's center, as shown in Fig. \ref{fig3}(c). Thus, its Fourier decomposition yields that the not pure odd harmonics should vanish, having very low values in practice. 
In Fig.~\ref{fig4}(c), the subtraction of these two terms is shown, which defines $S(k_x,k_y)$. The smoothness of the resulting function, with no discrimination between even or odd modes, indicates that the density fluctuations are governed by long wavelength dynamics and not by the static density profile.

The structure factor presents a notorious pre-peak centered at $kd \sim 0.2$, which corresponds to a large wavelength structure of size $\sim 15d$ (the pre-peak term refers to the peak being at lower wavenumber than the one corresponding to the first coordination shell at $kd = 2\pi$, see below). The associated density fluctuations are indeed visible by simple visual inspection (see Fig.~\ref{raw_img}). Fig. \ref{fig4}(d) shows $S(k)$ for a wider range of $kd$, up to $kd \sim 40$. It has the usual form expected for liquids with short range order, presenting a maximum at $kd = 2\pi$ related to the first coordination shell as well as the pre-peak discussed before. This pre-peak can be located in the range $kd = 0.1 - 0.3$ depending on the exact value of $\Gamma$. Similar density fluctuations have been observed in amorphous materials \cite{Elliott1991,Tanaka2005b}, which have been consistently related to the existence of medium-range-crystalline-order. These density fluctuations have also been related to the presence of a pre-peak in the structure factor. In fact, in the amorphous literature the pre-peak is known as the First Sharp Diffraction Peak (FSDP) because it appears at low $k$ and is obtained from X-ray diffraction measurements.

Fig. \ref{fig5}(a) presents $S(k)$ for small wavenumbers at different accelerations $\Gamma$ below $\Gamma_c$. From this figure it is clear how the pre-peak evolves as the system is driven towards the transition by increasing its acceleration towards $\Gamma_c$: its height increases and its position shifts to lower $k$. This implies that density fluctuations become larger in size and stronger as $\Gamma$ increases. We characterize the pre-peak by its maximum value $S_{\rm max} \equiv S(k^*)$, and the associated characteristic length scale $\xi = \pi/k^*$, where $k^*$ is the position of the pre-peak. These quantities are plotted in Figs. \ref{fig5}(b) and \ref{fig5}(c) as functions of $\Gamma$ for increasing amplitude ramps and for both experiment types A and B. Although the data points are scattered and that they do not really overlap, specially at higher $\Gamma$, both quantities show similar trends for each experimental type. Both $S_{\rm max} $ and $\xi/d$ increase as the transition is approached, although the former seems to saturate at larger $\Gamma$. We observe no great differences between the two ITO coatings, being their final values (near the transition) very similar, $S_{\rm max} \approx 0.6 -0.1$ and $\xi/d \approx 20-35$. We also present in Fig. \ref{fig5}(d) the width of the pre-peak, $\lambda$, defined as its width at half $S_{\rm max}$. This quantity is a measure of the dispersion around the characteristic length $\xi$. The collapse and scatter of the data are improved with respect to the other quantities, with no dependence on the different dissipation parameters.  

By observing visually the persistence of the solid clusters we conclude that for the A type experiment, the transition is located at $\Gamma_c \sim 5.1$, whereas for type B it is found to be $\Gamma_c \sim 4.5$. Given that we are dealing with a continuous phase transition these are just approximated values. However, neither $S_{\rm max}$ nor $\xi$ show evident changes at these values.

As a conclusion to this first part we can say that density fluctuations do not show critical behavior, but they are needed to create regions of high order. This is also evident from visual inspection; higher density patches are indeed more ordered, as can be verified in 
Fig.\ \ref{raw_img}. Density is a conserved field. Its fluctuations are however limited by the system's vertical geometrical constrain and the fact that the particles are in practice hard spheres. In what follows, medium range order will be analyzed with an appropriate bond-orientational order parameter, which does indeed present critical behavior. 

\subsection{Bond-orientational order parameter.}

\begin{figure}[t!]
\begin{center}
\includegraphics[width=0.6\columnwidth]{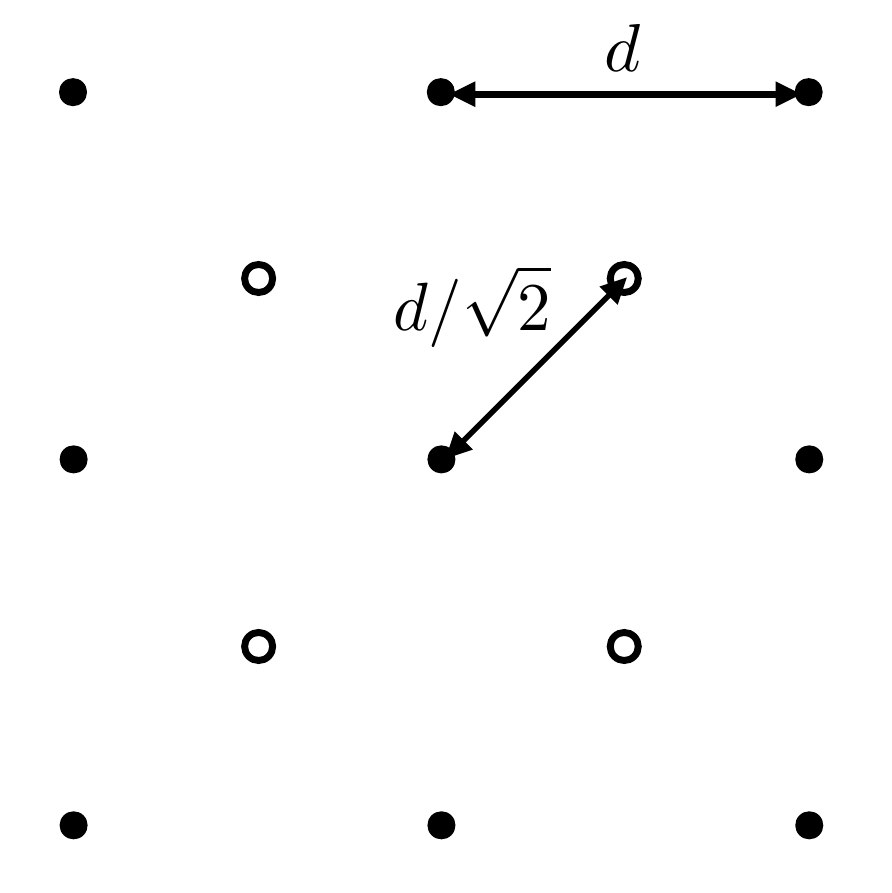}
\caption{Schematic representation of two square interlaced layers for which particles are closed packed. The center of particles in the bottom (top) layer are shown with solid (open) black circles. The 2D projected square lattice has a unit cell length $d/\sqrt{2}$.}
\label{Proyeccion2D}
\end{center}
\end{figure}

In the vicinity of the transition, fluctuations of high density present the same square symmetry as the solid phase. In the quasi-2D geometry the solid phase consists of two square interlaced layers instead of the hexagonal layer that is characteristic of 2D systems~\cite{Melby2005}. The local order can be characterized through a 4-fold bond-orientational order parameter. This is still valid in quasi-2D geometry because the interlaced two-layer square lattices (with unit cell length $d$ in each plane) result also in a square lattice when projected in 2D, with unit cell length $d/\sqrt{2}$ when the grains are close packed, as shown in Fig. \ref{Proyeccion2D}. The 4-fold bond-orientational order parameter per particle is defined
\begin{equation}
Q_4^j = \frac{1}{N_{j}} \sum_{s = 1}^{N_{j}} e^{4i\alpha_{s}^j},
\end{equation}
where $N_j$ is the number of nearest neighbors of particle $j$ and $\alpha_s^j$ is the angle between the neighbor $s$ of particle $j$ and the $x$ axis.  For a particle in a square lattice, $|Q_4^j | = 1$ and the complex phase measures  the square lattice orientation respect to the $x$ axis. For details on the computation of $Q_4^j$ we refer the reader to the supplementary information of Ref. \cite{castillo_PRL}. 

\begin{figure*}[t!]
\begin{center}
\includegraphics[width=0.97\columnwidth]{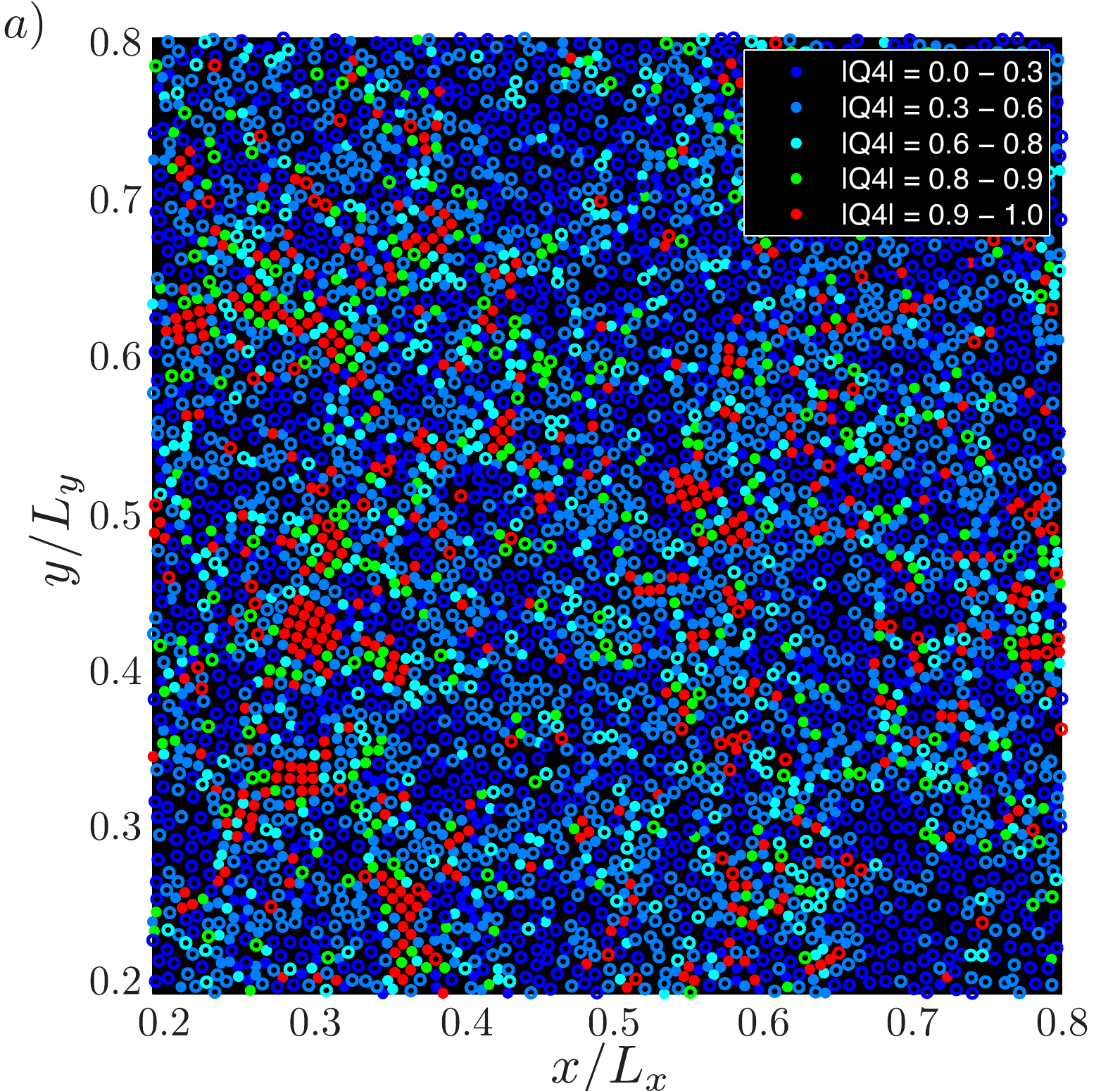}
\includegraphics[width=0.97\columnwidth]{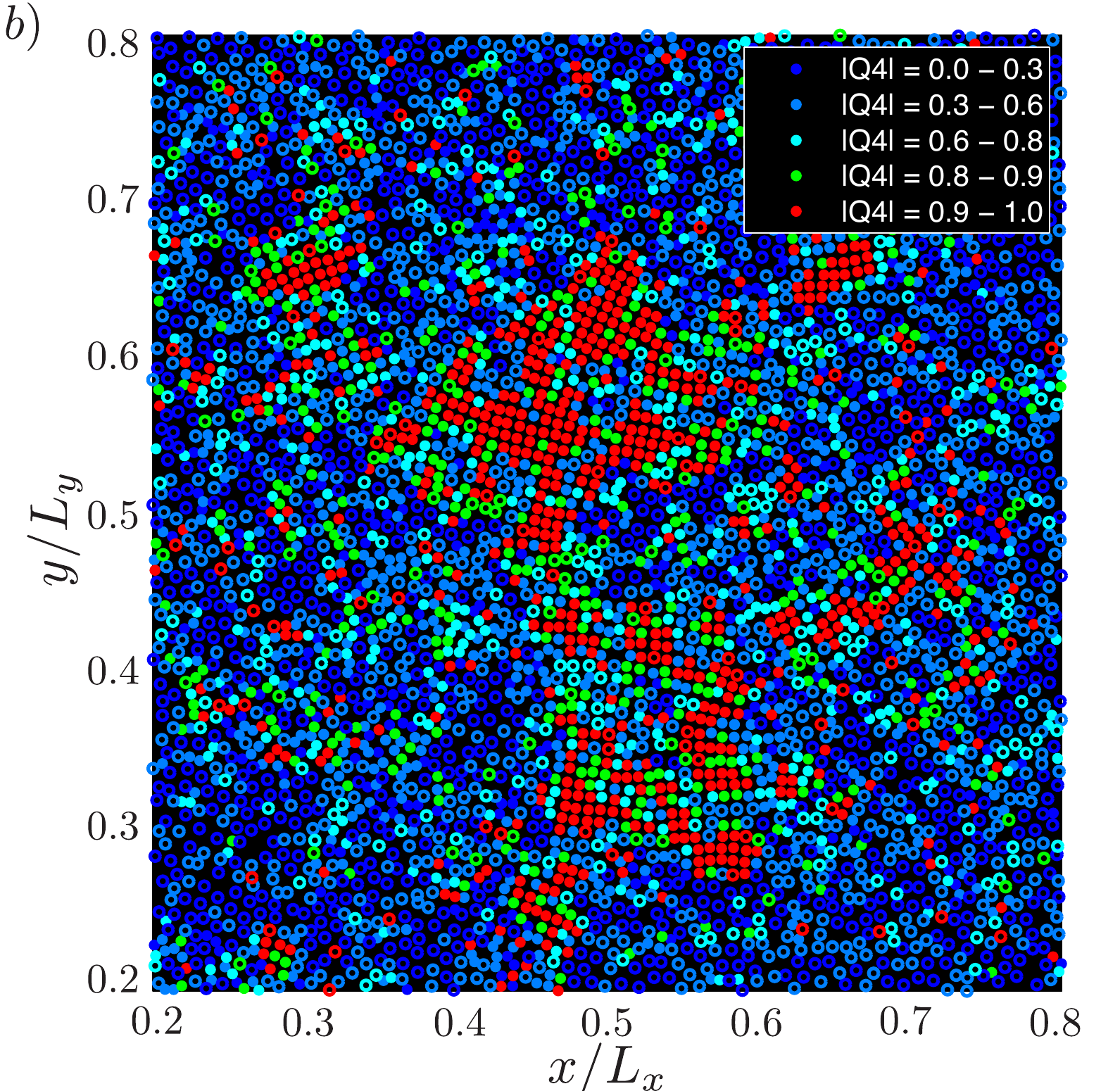}
\vspace{0.7cm}

\includegraphics[width=0.97\columnwidth]{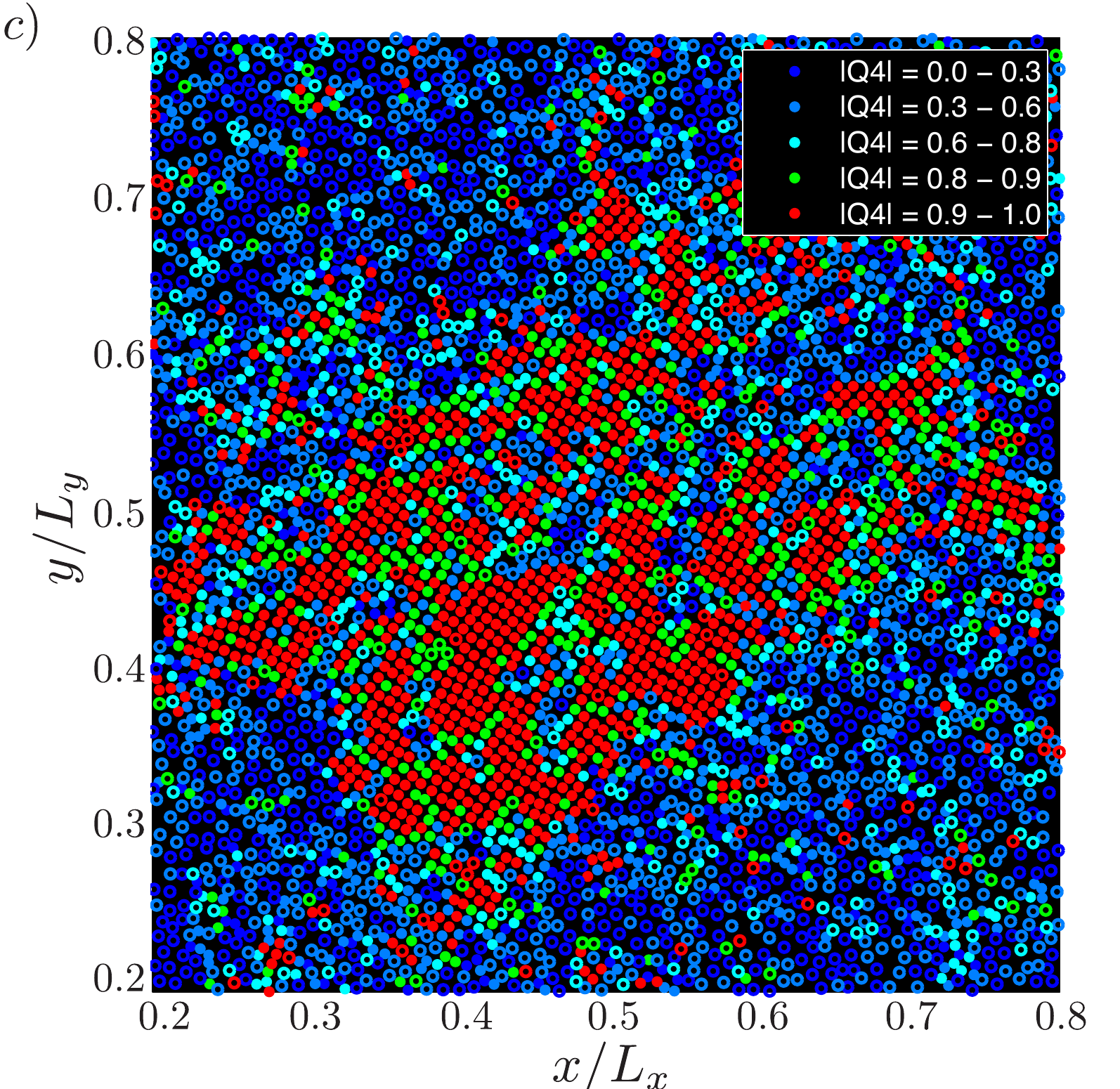}
\caption{(Color online) Color maps of the absolute value of the 4-fold bond-orientational order parameter in real space (for each particle we plot $|Q_4^j|$) for $\Gamma=4.18$ (a), $5.10$ (b) and $5.42$ (c) (Experiment type A, $\Gamma_c\approx 5.1$). In each figure only a part of the system is shown, from $0.2L$ to $0.8L$ in each horizontal dimension. The coloring is detailed in the legend, with the most ordered particles in red ($|Q_4| = 0.9 - 1$). Solid (open) circles correspond to particles classified as particles in the solid (liquid) phase. This classification is performed by measuring the area of the Voronoi area of each particle (see supplementary document of \cite{castillo_PRL}).}
\label{fig6}
\end{center}
\end{figure*}

\begin{figure*}[t!]
\begin{center}
\includegraphics[width=0.95\columnwidth]{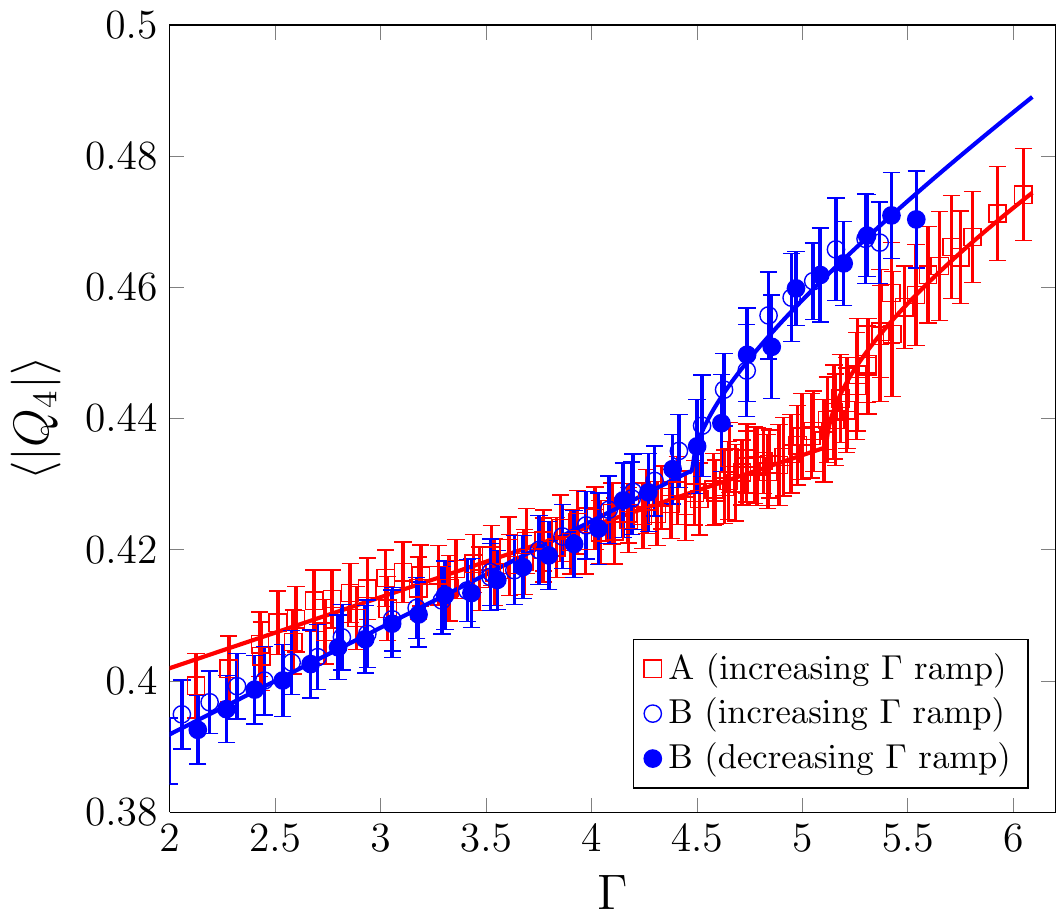}
\includegraphics[width=0.95\columnwidth]{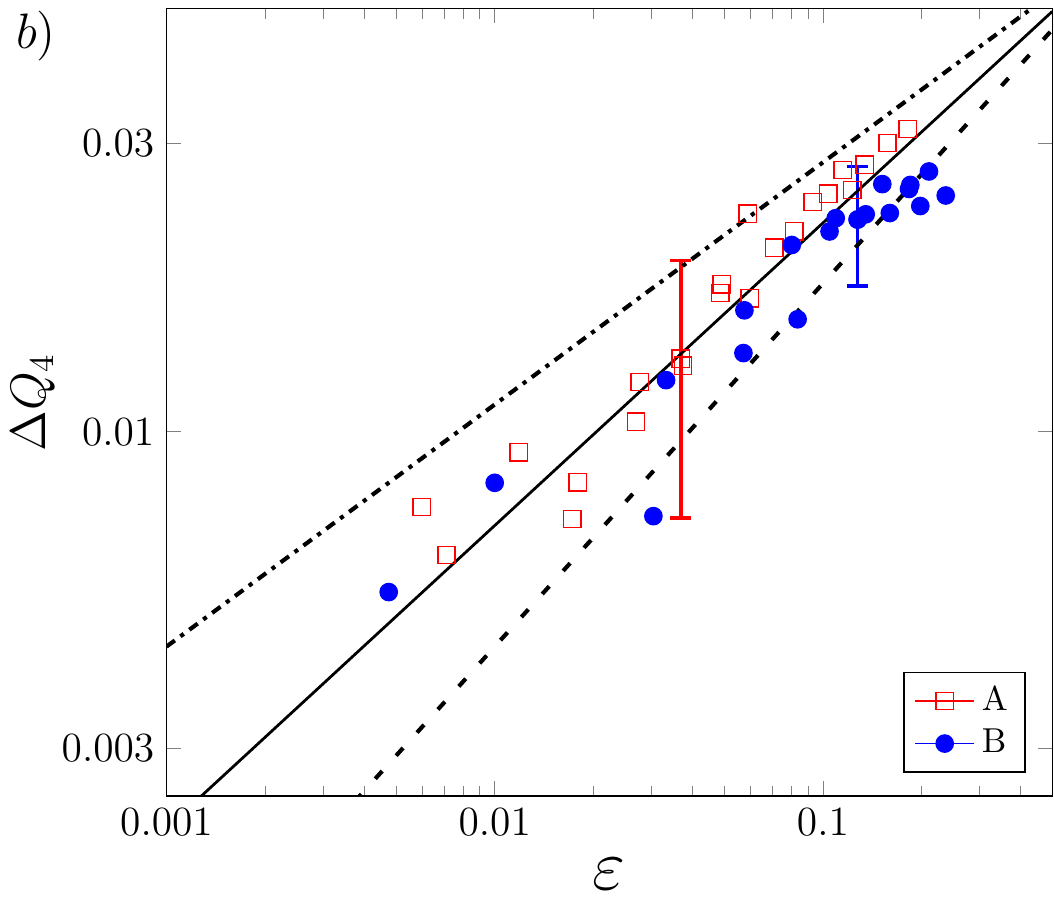}
\caption{(Color online) (a) Global average of 4-fold orientational order parameter $\langle |Q_4| \rangle$ versus $\Gamma$ for the two ITO coatings: Experiment type A (thin ITO) with increasing $\Gamma$ ramps ({$\square$}) and experiment type B (thick ITO) with increasing ($\circ$) and decreasing ($\bullet$) $\Gamma$ ramps. Continuous lines show the linear trend fit for $2.5<\Gamma<\Gamma_c$ and the supercritical deviation for $\Gamma \geqslant \Gamma_c$. 
The adjusted critical accelerations are $\Gamma_c = 5.12\pm0.01$ (type A) and $\Gamma_c = 4.48\pm0.03$ (type B).
(b) $\Delta Q_4 = \langle |Q_4| \rangle - Q_4^L$ versus $\varepsilon=(\Gamma-\Gamma_c)/\Gamma_c$ in log-log scale for each ITO coating thickness (type A: $\square$; type B $\circ$ for both increasing and decreasing $\Gamma$ ramps), where $Q_4^L = a\Gamma + b$ is obtained from the linear trend below $\Gamma_c$. For the thin ITO coating, $a = 0.011\pm0.001$ and $b=0.380\pm 0.002$. For the thick ITO coating, $a = 0.016\pm0.001$ and $b=0.359\pm 0.002$. In sake of clarity just one representative error bar is shown for each case.
The straight lines are power laws with exponents equal to $0.4$ (dash-dotted), $1/2$ (continuous) and $0.6$ (dashed), shown as guides to the eye. }
\label{Q4_global}
\end{center}
\end{figure*}

Representative maps of $|Q_4^j|$ are shown in Fig. \ref{fig6} for three accelerations, $\Gamma < \Gamma_c$, $\Gamma \approx \Gamma_c$ and $\Gamma > \Gamma_c$. In this case the maps are obtained from  images for experiment type A ($\Gamma_c\sim 5.1$). Below the transition the ordered patches, or crystallites, are first small, more or less homogeneously distributed in space and are of short lifetime. They increase in size and live for longer time as $\Gamma$ approaches $\Gamma_c$. Also, they tend appear more near the center than at the sidewalls, which we relate to the large scale small density inhomogeneity discussed before. The quantitative study of crystallites size and lifetime is presented below.

As discussed before, the local density can not change strongly at the transition because of the system's vertical geometrical  confinement. Thus, the correct order parameter must be the local 4-fold symmetry order, measured through the orientational parameter $Q_4^j$. With this in mind, we define its global average, in space first and then in time,
\begin{equation}
\langle |Q_4| \rangle = \left \langle \frac{1}{N} \sum \limits_{j=1}^{N} |Q_4^j| \right \rangle,
\end{equation}
which measures the fraction of particles in the ordered phase. 

\begin{figure}[t!]
\begin{center}
\includegraphics[width=0.93\columnwidth]{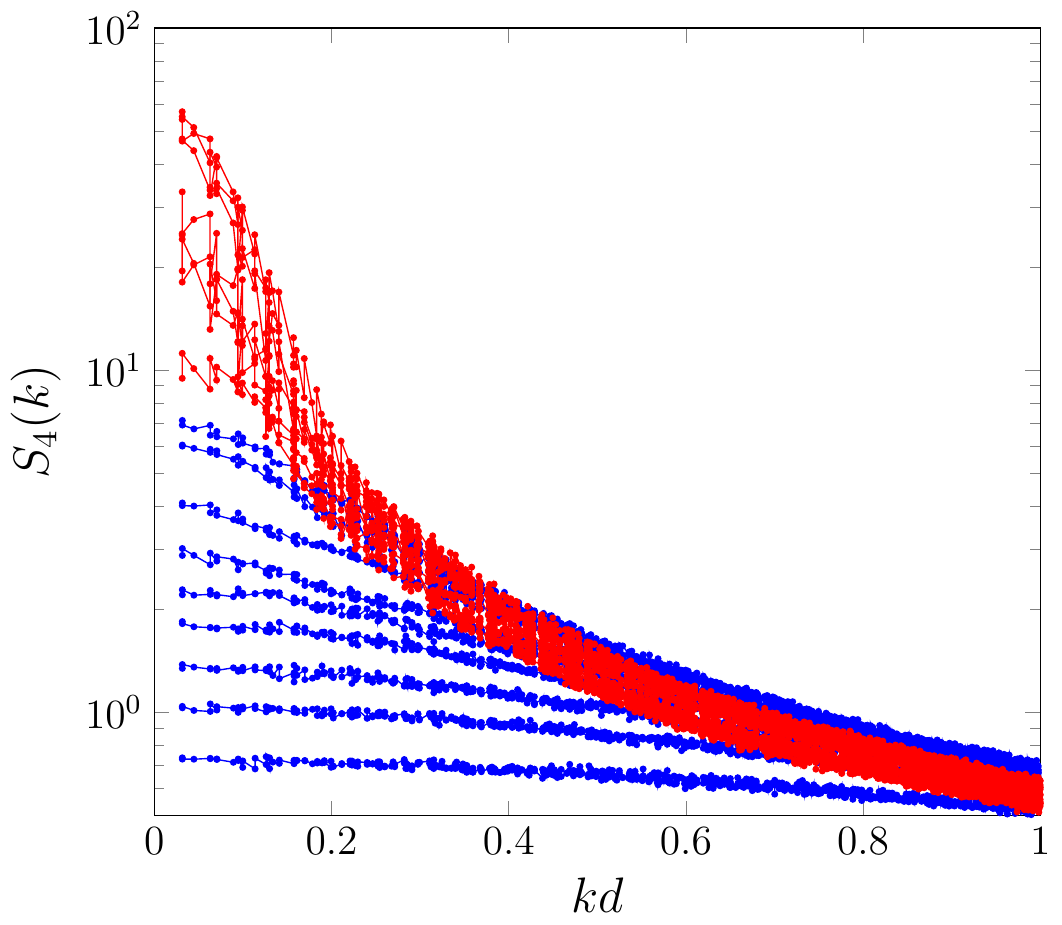}
\caption{(Color online) 4-fold bond-orientational structure factor $S_4(k)$ for several $\Gamma$ for experiment type A (the results for B are basically the same). The vertical axis is in $\log_{10}$ scale. Curves obtained for $\Gamma<\Gamma_c$ are in blue (dark gray) and for $\Gamma>\Gamma_c$ are in red (light gray). For both ITO thicknesses and for $\Gamma < \Gamma_c$, all curves show an Ornstein-Zernike-like behavior in the limit $kd\ll1$, $S_4(k) \approx S_4(0)/[1+(\xi_4 k)^2]$. For $\Gamma>\Gamma_c$ the curves tend to collapse together.}
\label{S4k}
\end{center}
\end{figure}

In order to show that the transition is indeed of second order for both experiment types, we present in Fig. \ref{Q4_global}a the global average $\langle |Q_4| \rangle $ versus $\Gamma$ for the thin and thicker ITO coated plates (type A and B respectively). For the latter, both increasing and decreasing $\Gamma$ ramps are presented, showing good reproducibility. This figure indeed demonstrates that both configurations present a second order type transition, continuous and with no hysteresis.

\begin{figure*}[t!]
\begin{center}
\includegraphics[width=\columnwidth]{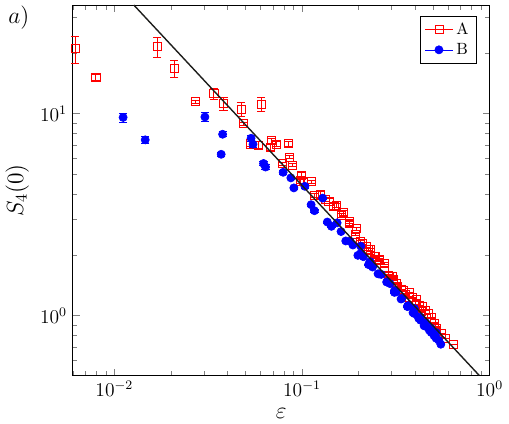}
\includegraphics[width=\columnwidth]{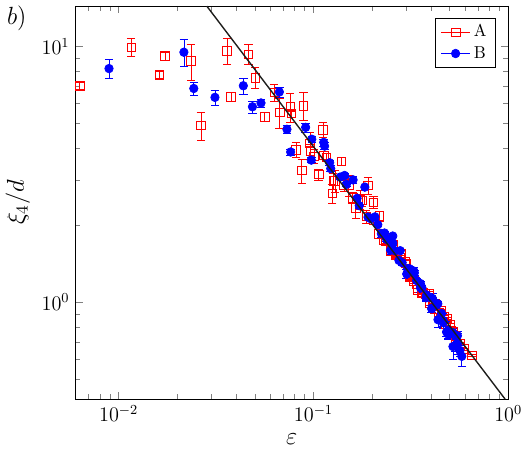}
\caption{(Color online) $S_4(0)$ (a) and $\xi_4/d$ (b) versus $\varepsilon$ for experiments type A ($\square$) and type B ($\circ$). Continuous lines are critical power laws, with exponents $\gamma = \nu_{\perp} = 1$ for $S_4(0)$ and $\xi_4$, shown as guides to the eye. The fitted critical accelerations for $S_4(0)$ and $\xi_4/d$ are: $\Gamma_c = 5.09\pm0.07$ and $\Gamma_c = 5.24\pm0.08$ respectively for experiment type A; $\Gamma_c = 4.43\pm0.06$ and $\Gamma_c = 4.58\pm0.06$ respectively for experiment type B. More details are provided in Table \ref{table1}. }
\label{S4-xi4}
\end{center}
\end{figure*}

The qualitative behavior is the same for both experiment types. First, $\langle |Q_4| \rangle $ has a linear dependence on $\Gamma$ below the transition. This reflects the fact that the fraction of particles that form crystallites with square fold symmetry, even if it is transiently, increases when the transition is approached. For both experiments A and B there is a clear deviation from this linear trend above a given threshold. The critical acceleration --defined for now as the value where the qualitative change occurs-- for the thicker ITO coating is lower ($\Gamma_c\sim4.5$) than the one for the thin ITO coating ($\Gamma_c \sim 5.1$). Also, the initial linear slope below the transition is larger for the thick ITO coating case. Both facts, the lower critical value and larger slope for the thicker ITO coating, are consistent with a lower effective dissipation at the top and bottom walls. Indeed, in this case the transition occurs at a lower amplitude, thus at lower energy injection and dissipation rates, and transient crystals form and grow more easily as $\Gamma$ increases. In this figure, the continuous lines correspond to fits of a linear dependence for  $\Gamma<\Gamma_c$ and a supercritical deviation for $\Gamma>\Gamma_c$ (details in the figure caption).

The deviation from the linear trend observed for $\Gamma<\Gamma_c$ is defined as $\Delta Q_4 = \langle |Q_4|\rangle- Q_4^L$, where $Q_4^L$ is defined as the extrapolation of the linear trend over the complete range of $\Gamma$. Fig. \ref{Q4_global}(b) presents $\Delta Q_4$ versus $\varepsilon=(\Gamma-\Gamma_c)/\Gamma_c$ in log-log scale for each ITO coating thickness. The continuous line shows the supercritical law $\Delta Q_4~\propto~ \sqrt{\Gamma-\Gamma_c}$ as a guide to the eye.

The results of Fig. \ref{Q4_global}(a) are fitted to the supercritical law $\Delta Q_4 = c\sqrt{\Gamma-\Gamma_c}$. The adjusted parameters are $c = 0.029\pm0.002$ (type A) and $c = 0.024\pm0.002$ (type B).
We conjecture that the different adjusted $c$ values also reflect the difference of dissipation parameters that control the particle-wall collisions.

Next, in order to characterize quantitatively the ordered patches shown in Fig. \ref{fig6}, in particular their typical length and time scales, we analyze the orientational order parameter in momentum space. Its Fourier components are 
\begin{equation}
{Q}_4(\vec k,t) = \sum_{j=1}^{N} Q_4^j  e^{i \vec k \cdot \vec r_j(t)}.
\end{equation}
Then, local order can also be analyzed through its fluctuations in Fourier space by means of the 4-fold bond-orientational structure factor
\begin{equation}
S_4(\vec k) = \frac{\langle | Q_4(\vec k,t) - \langle Q_4(\vec k,t) \rangle|^2\rangle}{N}. \label{defSk4}
\end{equation}

This structure factor is shown in Fig. \ref{S4k} for several accelerations and for experiment type A. Results for experiments type B are basically the same and are not shown. For both experiment types and for $\Gamma<\Gamma_c$, $S_4(k)$ shows an Ornstein-Zernike-like behavior in the limit $kd\ll1$,
\begin{equation}
S_4(k) \approx \frac{S_4(0)}{1+(\xi_4 k)^2},
\end{equation}
where $\xi_4$ and $S_4(0)$ are the 4-fold bond-orientational correlation length and static susceptibility respectively. 


\begin{table*}[t!]
\caption{Adjusted parameters with free or fixed critical exponents
for both ITO coatings. The fitted power laws are $\Delta Q_4 = c(\Gamma-\Gamma_c)^\beta$, $S_{4}\left(0\right) = \tilde a \varepsilon^{-\gamma}$, $\xi_{4}/d  = \tilde b \varepsilon^{-\nu_\perp}$, $\tau_{4}/T  = \tilde c \varepsilon^{-\nu_\parallel}$ and $S_4(k) = {C_\infty}/{k^{2-\eta}}$. The fit of the relaxation time $\tau_{4}$ corresponds to $kd=0.09$. For $\beta$ ($\eta$) only the fixed (free) case was analyzed.}
\begin{ruledtabular}
\begin{tabular}{ccccc}

\multicolumn{3}{l}{{Experiment type A -- Thin ITO coating -- larger wall dissipation}}\tabularnewline
\hline 
\multicolumn{5}{l}{Free critical exponents}\tabularnewline
& $\;\gamma=0.95\pm0.15\;$ & $\;\nu_{\perp}=0.97\pm0.21\;$ & $\;\nu_{\parallel}=1.98\pm0.50$  & $\eta = 0.69\pm0.01$ ($\Gamma = 5.04$) \tabularnewline
& $\Gamma_{c}=4.94\pm0.22$ & $\Gamma_{c}=5.11\pm0.35$ & $\Gamma_{c}=5.06\pm0.38$  & $\eta = 0.67\pm0.01$ ($\Gamma = 5.10$) \tabularnewline
& $\tilde{a}=0.47\pm0.03$ & $\tilde{b}=0.41\pm0.04$ & $\tilde{c}=1.11\pm0.13$  & $\eta = 0.59\pm0.01$ ($\Gamma = 5.12$)  \tabularnewline
\hline 
\multicolumn{5}{l}{Fixed critical exponents}\tabularnewline
$\beta = 1/2$ & $\gamma=1$ & $\nu_{\perp}=1$ & $\nu_{\parallel}=2$  & \tabularnewline
$\Gamma_c = 5.12\pm0.01$ & $\Gamma_{c}=5.09\pm0.07$ & $\Gamma_{c}=5.24\pm0.08$ & $\Gamma_{c}=5.12\pm0.07$  & \tabularnewline
$c = 0.029 \pm 0.002$ & $\tilde{a}=0.47\pm0.01$ & $\tilde{b}=0.41\pm0.01$ & $\tilde{c}=1.10\pm0.03$  & \tabularnewline
\hline
\multicolumn{3}{l}{{Experiment type B -- Thick ITO coating -- lower wall dissipation}}\tabularnewline
\hline 
\multicolumn{5}{l}{Free critical exponents}\tabularnewline
&$\gamma=1.10\pm0.16$ & $\nu_{\perp}=1.10\pm0.03$ & $\nu_{\parallel}=1.98\pm0.21$ & $\eta = 0.87\pm0.01$ ($\Gamma = 4.29$) \tabularnewline
&$\Gamma_{c}=4.54\pm0.18$ & $\Gamma_{c}=4.73\pm0.03$ & $\Gamma_{c}=4.55\pm0.18$ & $\eta = 0.67\pm0.01$ ($\Gamma = 4.41$)\tabularnewline
&$\tilde{a}=0.39\pm0.03$ & $\tilde{b}=0.35\pm0.03$ & $\tilde{c}=1.14\pm0.07$ & $\eta = 0.63\pm0.01$ ($\Gamma = 4.52$)\tabularnewline
\hline 
\multicolumn{5}{l}{Fixed critical exponents}\tabularnewline
$\beta = 1/2$ & $\gamma=1$ & $\nu_{\perp}=1$ & $\nu_{\parallel}=2$ & \tabularnewline
$\Gamma_c = 4.48\pm 0.03$ &$\Gamma_{c}=4.43\pm0.06$ & $\Gamma_{c}=4.58\pm0.06$ & $\Gamma_{c}=4.46\pm0.03$ & \tabularnewline
$c = 0.024 \pm 0.002$ & $\tilde{a}=0.41\pm0.01$ & $\tilde{b}=0.37\pm0.01$ & $\tilde{c}=1.17\pm0.01$ & \tabularnewline
\end{tabular}
\end{ruledtabular}

\label{table1}
\end{table*}

In Fig. \ref{S4-xi4} we present both $S_4(0)$ and $\xi_4/d$ for $\Gamma < \Gamma_c$, obtained from the fits of the Ornstein-Zernike behavior for both ITO coatings. Both quantities are plotted as functions of the reduced acceleration $\varepsilon = (\Gamma_c - \Gamma)/\Gamma_c$, where $\Gamma_c$ is obtained from a specially adapted fitting procedure \cite{castillo_PRL}. The correlation length and susceptibility vary strongly as the transition is approached. For the two ITO coatings these quantities obey critical power laws. In the limit $\varepsilon \rightarrow 0$ they both saturate, presumably due to the system's finite size. For $\varepsilon \lesssim 3\times10^{-2}$ they saturate to $S_4(0)\approx 10-20$ and $\xi_4/d\approx 10$ respectively. The critical-like behavior is fitted with both free and fixed exponents. As discussed in our previous work \cite{castillo_PRL}, the precise measurement of $\Gamma_c$ and the critical exponents is far from trivial. Initial fits give $\Gamma_c$ in the range $5.1-5.6$ and exponents that can vary up to a factor of almost 2. Additionally, the fitted $\Gamma_c$ can be quite different depending from which quantity they are obtained. The lack of precision is due to the arbitrariness in the choice of the range $\Gamma$ to be used for the fit. In \cite{castillo_PRL} we present a robust method for the determination of these quantities, readers are referred to its supplementary information document for details. 

The results for both experiment types are given in Table \ref{table1} (the columns referring to the relaxation time $\tau_4$ and $S_4(k)$ in the hydrodynamic regime will be discussed below). We conclude that for both ITO coatings, the exponents are the same within experimental errors.

In what follows, the exponents are assumed to be fixed and we define 
\begin{equation}
S_4(0) = \tilde a \varepsilon^{-\gamma},\quad\quad\quad
\xi_4/d = \tilde b \varepsilon^{-\nu_{\perp}},
\end{equation}
with the critical exponents $\gamma=1$ and $\nu_{\perp}=1$. The critical divergence with $\varepsilon$ makes it necessary to fit the $\Gamma_c$ separately for each case. For experiment type A the adjusted critical accelerations are $\Gamma_c=5.09~\pm~0.07$ and $\Gamma_c = 5.24\pm0.08$ for $S_4(0)$ and $\xi_4$ respectively, whereas for B they are $\Gamma_c= 4.43\pm0.06$ and $\Gamma_c = 4.58\pm0.06$ respectively. For each ITO coating we find that within experimental errors both critical accelerations are very consistent, as well as with the value obtained from the supercritical behavior of $\Delta Q_4$ ($\Gamma_c = 5.12\pm0.01$ and $\Gamma_c = 4.48\pm0.03$ for A and B respectively). Notice that now the critical accelerations are obtained from fits of measured quantities below the transition, whereas for $\Delta Q_4$ they were obtained with fits of the order parameter above the transition. Finally, the critical accelerations are less consistent between the different measured quantities when the critical exponents are let to be free parameters. This supports the choice of fixing the critical exponents.

\begin{figure}[t!]
\begin{center}
\includegraphics[width=0.93\columnwidth]{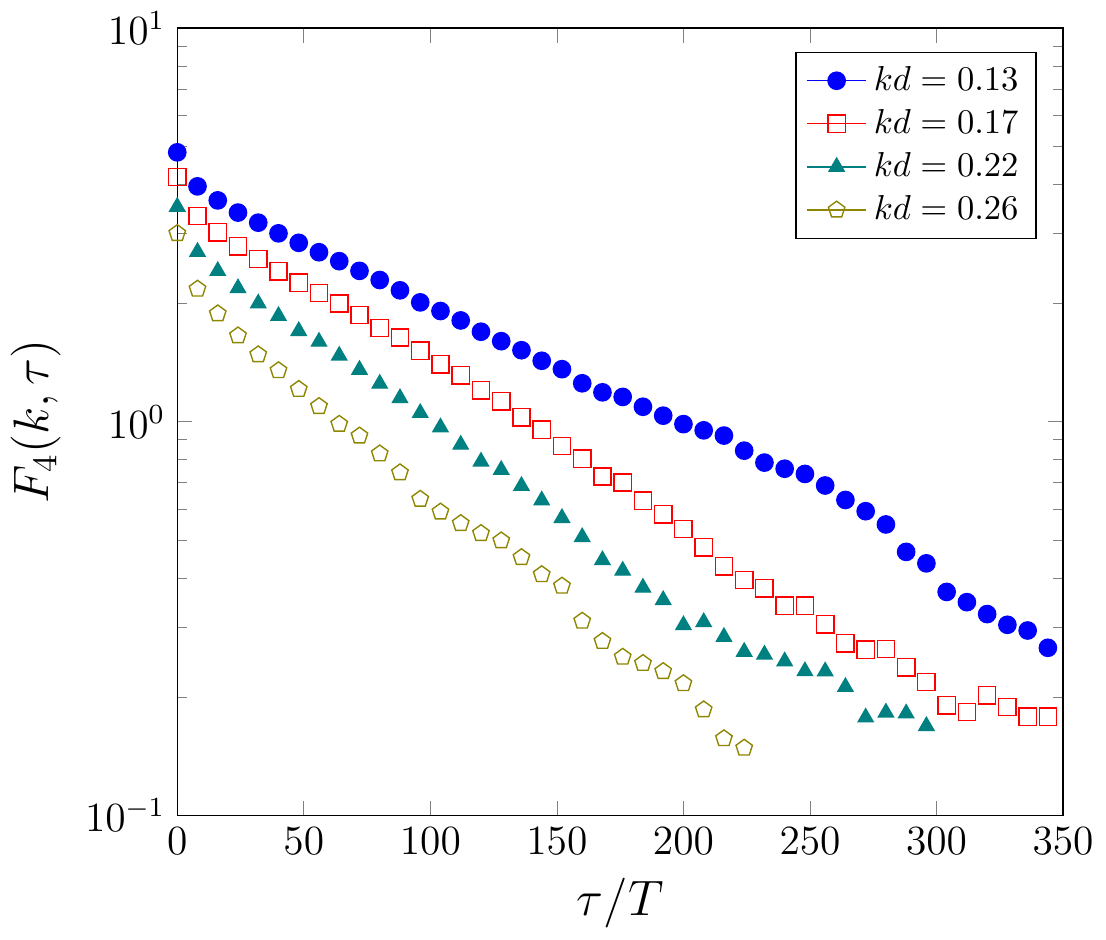}
\caption{(Color online) Dynamic 4-fold bond-orientational structure factor $F_4(k,\tau)$ for several $kd$ and $\Gamma=4.36<\Gamma_c$ (experiment type B), which shows an exponential decay. Large wavelengths (small wavenumbers) decay slower than short wavelengths (large wavenumbers). The behavior of $F_4(k,\tau)$ is basically the same for both ITO coatings. Only data above de noise level is shown for each $kd$.}
\label{F4_new}
\end{center}
\end{figure}

\begin{figure*}[t!]
\begin{center}
\includegraphics[width=\columnwidth]{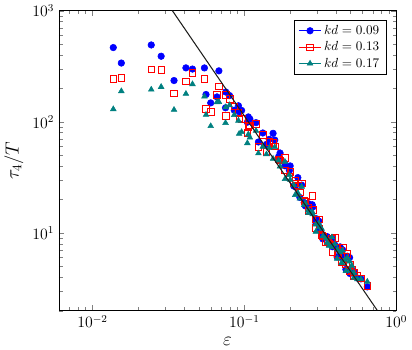}
\includegraphics[width=\columnwidth]{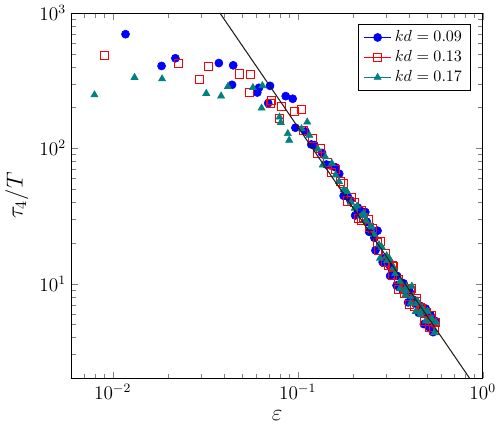}
\caption{(Color online) $\tau_4/T$ versus $\varepsilon$ for experiment type A (left) and B (right) for three wavenumbers. The continuous lines are critical power laws with exponents equal to $-2$, shown as guides to the eye. The fitted critical accelerations are $\Gamma_c = 5.12\pm0.07$ and $\Gamma_c = 4.46\pm0.03$ for A and B respectively. }
\label{tau4}
\end{center}
\end{figure*}

Now, we present the characterization of the crystallite's relaxation time. Time correlations are computed through the two-time bond-orientational  correlation function
\begin{equation}
F_4(\vec k,\tau) = \frac{\langle \delta Q_4(\vec k,t+\tau) \delta  Q_4(\vec k,t)^* \rangle}{N}, \label{defSk4tau}
\end{equation}
where $^*$ stands for the complex conjugate and
$\delta Q_4(\vec k,t) =  Q_4(\vec k,t) - \langle  Q_4(\vec k,t) \rangle$.
Our results show that for low wavevectors 
\begin{equation}
F_4(\vec k,\tau) \approx F_4(\vec k,0) \exp(-\tau/\tau_4(k)),
\end{equation}
from which the relaxation time $\tau_4(k)$ is measured. The exponential decay behavior is shown in Fig. \ref{F4_new}. From the measured relaxation times we also obtain a critical-like behavior, which is presented in Fig. \ref{tau4} for the particular cases $kd = 0.09$, $0.13$ and $0.17$. Here, again we obtain critical power law forms, for both ITO coatings. We have also used free and fixed critical exponents, for which details are given in Table \ref{table1}. The conclusion is the same as for the correlation length and susceptibility, namely that within experimental errors the free exponent adjustment is very consistent with a fixed critical exponent $\nu_\parallel = 2$. Thus, we assume this critical exponent to be fixed and we define 
\begin{equation}
\tau_4/T = \tilde c \varepsilon^{-\nu_{\parallel}},
\end{equation}
with $\nu_{\parallel}=2$ and $T$ is the vibration period. The adjusted critical accelerations are $\Gamma_c = 5.12\pm0.07$ and $\Gamma_c = 4.46\pm0.03$, for A and B respectively. The relaxation time also seems to saturate for small $\varepsilon$, which occurs at smaller $\varepsilon$ for lower $k$, that is for fluctuations of larger size. 

From the relaxation time analysis we conclude that the dynamical exponent is $z = \nu_{\parallel}/\nu_{\perp}=2$. As usual, there is critical slowing down in the dynamics. As a consequence, close to the critical point, stationary states are obtained after a long relaxation has taken place. Taken that into account, all $\Gamma$ ramps are chosen to be slow. Also, averages are taken for long times.

As a final evidence of the observed criticality we now turn to the characterization of the anomalous exponent~$\eta$. In the hydrodynamic regime, $d/\xi_4 \ll kd \ll 1$, $S_4(k)$ is expected to present a power law decay 
\begin{equation}
S_4(k) \approx \frac{C_\infty}{k^{2-\eta}},
\end{equation}
where $\eta$ is the critical exponent related to the decay of the pair correlation function $g(r) \sim r^{D-2+\eta}$, with $D$ the dimensionality. 

\begin{figure*}[t!]
\begin{center}
\includegraphics[width=\columnwidth]{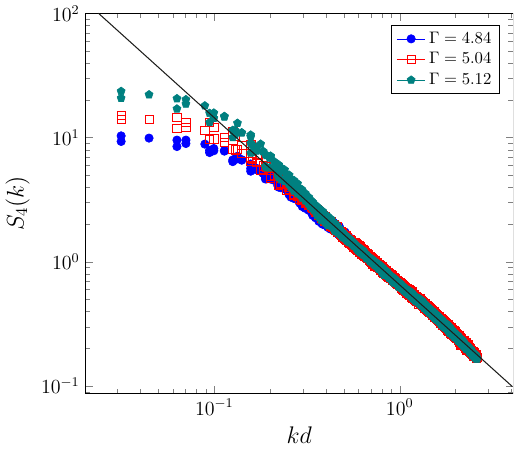}
\includegraphics[width=\columnwidth]{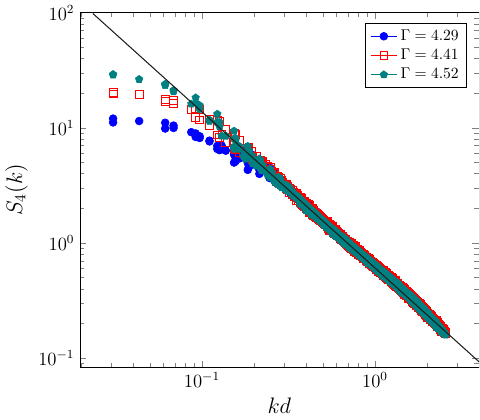}
\caption{(Color online) $S_4(k)$ in log-log scale for several $\Gamma$ for experiments type A (left) and type B (right). The continuous line corresponds to $\eta=0.63$. We recall that $\Gamma_c \approx 5.1$ and $\Gamma_c \approx 4.5$ respectively. As expected, the hydrodynamic regime becomes wider as the transition is approached and $S_4(k)$ obeys a power law. }
\label{etaB}
\end{center}
\end{figure*}

Fig. \ref{etaB} presents $S_4(k)$ in log-log scale for various $\Gamma$. Indeed, as the transition is approached, curves tend to collapse for shorter wavelengths. They are clearly different for larger wavelengths as they converge to different static susceptibilities $S_4(0)$. In principle, $\eta$ must be obtained in the limit $\Gamma\rightarrow \Gamma_c$. As the critical acceleration is not known with sufficient precision, in Fig. \ref{etaB} we present $S_4(k)$ for three accelerations in the vecinity of $\Gamma_c$ for both ITO coatings. For the highest $\Gamma$, it is not even certain that the system is below the transition or not. Performing power law fits constraining $k$ to the range $d/\xi_4 \leqslant kd  \leqslant 1$, for configuration A, the measured critical exponents are $\eta = 0.69 \pm 0.01$, $\eta = 0.67\pm0.01$ and $\eta = 0.59 \pm 0.01$ for $\Gamma = 5.04$, $5.10$ and $5.12$ respectively. Similarly, for experiment type B they are $\eta = 0.87 \pm 0.01$, $\eta = 0.67 \pm 0.01$ and $\eta = 0.63 \pm 0.01$ for $\Gamma = 4.29$, $4.41$ and $4.52$ respectively. Thus, although the anomalous exponent $\eta$ varies rather strongly depending on $\Gamma_c-\Gamma$, we can state that close enough to $\Gamma_c$ it can be bounded in the range $0.6 - 0.67$ for both ITO coatings. In conclusion, there is clearly a hydrodynamic regime for which the power behavior is valid, even for a wider range than predicted. However, the measurement of $\eta$ needs to be done extremely close to $\Gamma_c$. With the present data, we can state that $\eta \approx 0.6-0.67$ is a good estimation. 

\section{Discussion and conclusions} \label{sec.conclusions} 

We studied  the liquid-solid-like transition that takes place in confined quasi two dimensional granular systems. Two configurations that differ in the dissipation were analyzed, presenting the transition at different critical accelerations. We have demonstrated that the non-equilibrium  transition is a second order type for both configurations studied. 

Besides, we showed that in our experiments, for both cases, density fluctuations do not present strong variations at the transition. In fact, the static structure factor $S(k)$ actually presents a peak at low wavenumbers, which is related to the existence of medium range crystalline order \cite{Elliott1991}. In our case, the characteristic length $\xi$ of the these structures in the system does not show critical behavior.

On the contrary, local order presents critical behavior.
It is characterized through the bond-orientational order parameter $Q_{4}$, which in Fourier space shows an Ornstein-Zernike-like behavior. The associated correlation length $\xi_{4}$, the relaxation time $\tau_{4}$, the zero $k$ limit of $Q_{4}$ fluctuations (static susceptibility), the pair correlation function of $Q_{4}$, and the amplitude of the order parameter obey critical power laws, with saturations due to finite size effects. Their respective critical exponents are $\nu_{\perp}=1$, $\nu_{\parallel}=2$, $\gamma=1$, $\eta\approx 0.6 - 0.67$ and $\beta=1/2$, whereas the dynamical critical exponent $z=\nu_{\parallel}/\nu_{\perp}=2$. Although the critical accelerations and the pre-factors of the power laws differ between the two setups, the reported critical exponents are the same. Hence, while dissipation is strictly necessary to form the crystal, the path the system undergoes towards the phase separation is part of a well defined universality class.

In equilibrium, the scaling hypothesis predicts relations among the critical exponents. It is worth mentioning that the relation $\gamma=(2-\eta)\nu_{\perp}$ is not satisfied, while $\alpha+2\beta+\gamma=2$ and  $\nu_{\perp} D=2-\alpha$ ($D=2$ is the spatial dimension) can be satisfied simultaneously if $\alpha=0$. This exponent, associated in equilibrium to the specific heat divergence, has no interpretation out of equilibrium. 

The critical order parameter in the present case is a non-conserved complex scalar field. Its dynamics, however, is not expected to be autonomous even close to the critical point as density fluctuations are needed to create the ordered phase. Although it has been shown that the transition dynamics is mediated by waves~\cite{clerc2008}, momentum density decays fast due to friction. Therefore, the most appropriate description in the theory of dynamical critical phenomena is model C, in which a non-conserved order parameter is coupled to a conserved non-critical density~\cite{hohenberg}. In this case~\cite{hohenberg, Ccorrect}, and in extensions to non-equilibrium dynamics~\cite{noneqC}, the dynamical exponent is predicted to be $z=2+\alpha/\nu_{\perp}$, consistent with the measurements if $\alpha=0$.

\acknowledgments

We thank D. Risso and J. Silva for valuable technical help and discussions.  This research is supported by Fondecyt Grants No. 1120211 (N.M) and No. 1140778 (R.S.). G.C. acknowledges the financial support of Conicyt Postdoctorado-2013 74140075.

\end{document}